\documentclass[sigconf, screen, table]{acmart}

\usepackage{amsmath,amssymb,amsfonts}

\usepackage{textcomp}
\usepackage{graphicx}
\usepackage{booktabs}
\usepackage{caption}
\usepackage{subcaption}

\usepackage{multicol}
\usepackage{xcolor}

\usepackage{algorithm}
\usepackage{algpseudocode}
\usepackage{multirow,makecell}
\usepackage{units}

\usepackage{enumitem}

\algrenewcommand\algorithmicrequire{\textbf{Input:}}
\algrenewcommand\algorithmicensure{\textbf{Output:}}

\newcommand{\mycc}{\cellcolor{gray!15}}

\definecolor{mylightblue}{HTML}{9FA8DA}

\setcopyright{none}
\renewcommand\footnotetextcopyrightpermission[1]{}
\begin{document}

\title[\textit{Serinv}: A Scalable Library for BTA Matrices]{\textit{Serinv}: A Scalable Library for the Selected Inversion of \newline Block-Tridiagonal with Arrowhead Matrices}

\author{Vincent Maillou}
\orcid{0000-0003-4861-3298}
\affiliation{%
  \institution{D-ITET, ETH Zurich}
  \country{Switzerland}
}
\email{vmaillou@iis.ee.ethz.ch}

\author{Lisa Gaedke-Merzhaeuser}
\orcid{0000-0002-7586-2727}
\affiliation{%
  \institution{Institute of Computing, USI, CH}
  \country{}
}
\affiliation{%
  \institution{CEMSE Division, Kaust, SA}
  \country{}
}
\email{lisa.gaedkemerzhauser@kaust.edu.sa}

\author{Alexandros Nikolaos Ziogas}
\orcid{0000-0002-4328-9751}
\affiliation{%
  \institution{D-ITET, ETH Zurich}
  % \city{Zurich}
  \country{Switzerland}
}
\email{alexandros.ziogas@iis.ee.ethz.ch}

\author{Olaf Schenk}
\orcid{0000-0001-8636-1023}
\affiliation{%
  \institution{Institute of Computing, USI}
  % \city{Lugano}
  \country{Switzerland}
}
\email{olaf.schenk@usi.ch}

\author{Mathieu Luisier}
\orcid{0000-0002-2212-7972}
\affiliation{%
  \institution{D-ITET, ETH Zurich}
  % \city{Zurich}
  \country{Switzerland}
}
\email{mluisier@iis.ee.ethz.ch}

\begin{abstract} % 0.5 column
The inversion of structured sparse matrices is a key but computationally and memory-intensive operation in many scientific applications.
There are cases, however, where only particular entries of the full inverse are required.
This has motivated the development of so-called \textit{selected-inversion} algorithms, capable of computing only specific elements of the full inverse.
Currently, most of them are either shared-memory codes or limited to CPU implementations.
Here, we introduce \textit{Serinv}, a scalable library providing distributed, GPU-based algorithms for the selected inversion and Cholesky decomposition of positive-definite, block-tridiagonal arrowhead matrices. 
This matrix class is highly relevant in statistical climate modeling and materials science applications.
The performance of \textit{Serinv} is demonstrated on synthetic and real datasets from statistical air temperature prediction models.
In our numerical tests, \textit{Serinv} achieves 32.3\% strong and 47.2\% weak scaling efficiency and up to two orders of magnitude speedup over the sparse direct solvers PARDISO and MUMPS on 16 GPUs.
\end{abstract}




\maketitle

\pagestyle{plain}

\section{Introduction}\label{sec:introduction} % 3 columns

Many applications in materials science, computational chemistry, or statistical modeling require inverting large sparse matrices~\cite{negf1,omen,zammit2018sparse, gaedke2022parallelized} with a structured pattern and extracting selected entries of their inverse, for example, the diagonal elements.
Since matrix inversions generally lead to dense matrices, depending on the size of the problem at hand, such operations rapidly become unfeasible, either because of memory or timing constraints.
When specific entries of the inverse are needed, it is advantageous to compute only the desired ones without considering the others.
To achieve this, so-called \emph{selected-inversion} algorithms~\cite{takahashi_sellinv, erisman1975computing} have been derived to directly and exactly compute the desired elements of the inverse, thus speeding up the process and reducing the memory footprint.

\begin{table}[t]
\centering
\resizebox{\columnwidth}{!}{%
\begin{tabular}{ccccc}
\toprule
\multicolumn{2}{c}{\textbf{HW/Sparsity}} & \textbf{Unstructured} & \textbf{BT} & \textbf{BTA}\\
\midrule
\multirowcell{5}{\textbf{Shared}\\\textbf{Memory}} & \multirow{4}{*}{CPU} & \multirowcell{4}{$\text{MUMPS}^{\dagger}$~\cite{mumps_1, mumps_2, mumps_3}\\
$\text{Pardiso}^{*}$~\cite{pardiso_1, pardiso_2, pardiso_3, pardiso_4}\\
$\text{PSelInv}^{\dagger}$~\cite{lin_selinv, lin2009fast}\\
FIND~\cite{find_1, find_2, find_3, find_4}} & \multirowcell{4}{$\text{RGF}^{\dagger}$~\cite{rgf_1, rgf_2}} & \multirowcell{4}{$\text{INLA}_\text{BTA}^{\dagger}$~\cite{gaedkeIntegrated2024,inla_dist}}\\
&&&&\\
&&&&\\
&&&&\\
\cmidrule{2-5} & GPU & &$\text{RGF}^{\dagger}$~\cite{rgf_gpu}& $\text{INLA}_\text{BTA}^{\dagger}$~\cite{gaedkeIntegrated2024,inla_dist}\\
\midrule
\multirowcell{7}{\textbf{Distributed}\\\textbf{Memory}} & \multirow{4}{*}{CPU} & \multirowcell{4}{$\text{MUMPS}^{\dagger}$~\cite{mumps_1, mumps_2, mumps_3}}& \multirowcell{4}{\cellcolor[HTML]{00D2CB}$\text{Serinv}^{\dagger}$\\
PSR~\cite{petersen_hybrid_2009}\\
P-DIV/Spikes~\cite{pdiv_1, pdiv_2, pdiv_3, pdiv_4}\\
BCR~\cite{bcr_1, bcr_2}}& \multirowcell{4}{\cellcolor[HTML]{00D2CB}$\text{Serinv}^{\dagger}$}\\
&&&&\\
&&&&\\
&&&&\\
\cmidrule{2-5} & \multirow{3}{*}{GPU} & & \multirowcell{3}{\cellcolor[HTML]{00D2CB}$\text{Serinv}^{\dagger}$\\
SplitSolve~\cite{splitsolve}\\
P-DIV/Spikes~\cite{pdiv_gpu}} & \multirowcell{3}{\cellcolor[HTML]{00D2CB}$\text{Serinv}^{\dagger}$}\\
&&&&\\
&&&&\\
\bottomrule
\end{tabular}
}
\vspace{2pt}
\caption{Summary of existing selected-inversion algorithms, sorted by hardware (shared- and distributed-memory, CPUs, and GPUs) and sparse matrices type support (unstructured, BT, and BTA). We denote with a $\dagger$ ($*$) the open-source (closed-license) packages. Algorithms without annotation do not have any publicly available implementation to our knowledge. This work's contribution, \textit{Serinv}, is highlighted in cyan.}
\label{tab:litterature_position}
\end{table}

Selected inversion lends itself particularly well to Bayesian inference problems relying on integrated nested Laplace approximations (INLA) \cite{rue2009approximate}. The latter provides approximate inference estimates for latent Gaussian models (LGMs).
To save memory, such models can be described using sparse precision matrices, which typically possess well-structured sparsity patterns.
Block tridiagonal (BT) and block tridiagonal arrowhead (BTA) structures are very common in INLA~\cite{lindgren2024diffusion}.
To estimate the required marginal variances of latent parameters, selected entries of the precision matrix's inverse, i.e., its covariance matrix, must be extracted.

Selected-inversion algorithms rely on an appropriate decomposition of the matrix of interest, for example, via LU or Cholesky factorization. In the case of sparse matrices, the fill-in caused by these operations, i.e., zero entries of the matrix overwritten by non-zeros elements, represents a significant challenge as it increases the overall memory footprint and computational cost. 
This is particularly true for \textit{unstructured} matrices, where the coordinates of the non-zero entries do not follow any particular pattern.

Therefore, the decomposition and selected inversion of unstructured matrices necessitate algorithms oblivious to the distribution of the non-zero elements.
In matrices with a \textit{structured} (predefined)  sparsity pattern, the locations where non-zeros may appear during the decomposition phase are limited, providing unique opportunities for optimization~\cite{symbolic_cholesky}.

Of particular interest are structured matrices whose decomposition induced fill-in remains within the bound of their sparsity pattern. 

The aforementioned BT and BTA matrices fulfill this property.
They naturally appear in the quantum transport simulations of nanoscale materials and devices~\cite{negf1,omen} or INLA-based statistical climate modeling problems~\cite{gaedkeIntegrated2024}.
Alternatively, they can be constructed from unstructured matrices through decomposition and permutation~\cite{lukas,arrowhead_decomp}. 
By taking advantage of the BT and BTA sparsity patterns, the computational cost and memory footprint of selected-inversion and decomposition methods can be greatly reduced, typically from $O(N^3)$ down to $O(n \times b^3)$~\cite{rgf_gpu,rgf_2}, where $N=n \times b$ is the original system size, while $n$ is the number of diagonal blocks and $b$ their size.

%   - BTA parametrization (1 ~figure)
\begin{figure}[t]
    \centering
    % \begin{subfigure}[t]{0.72\columnwidth}
    \begin{subfigure}[t]{0.84\columnwidth}
        \includegraphics[width=\textwidth]{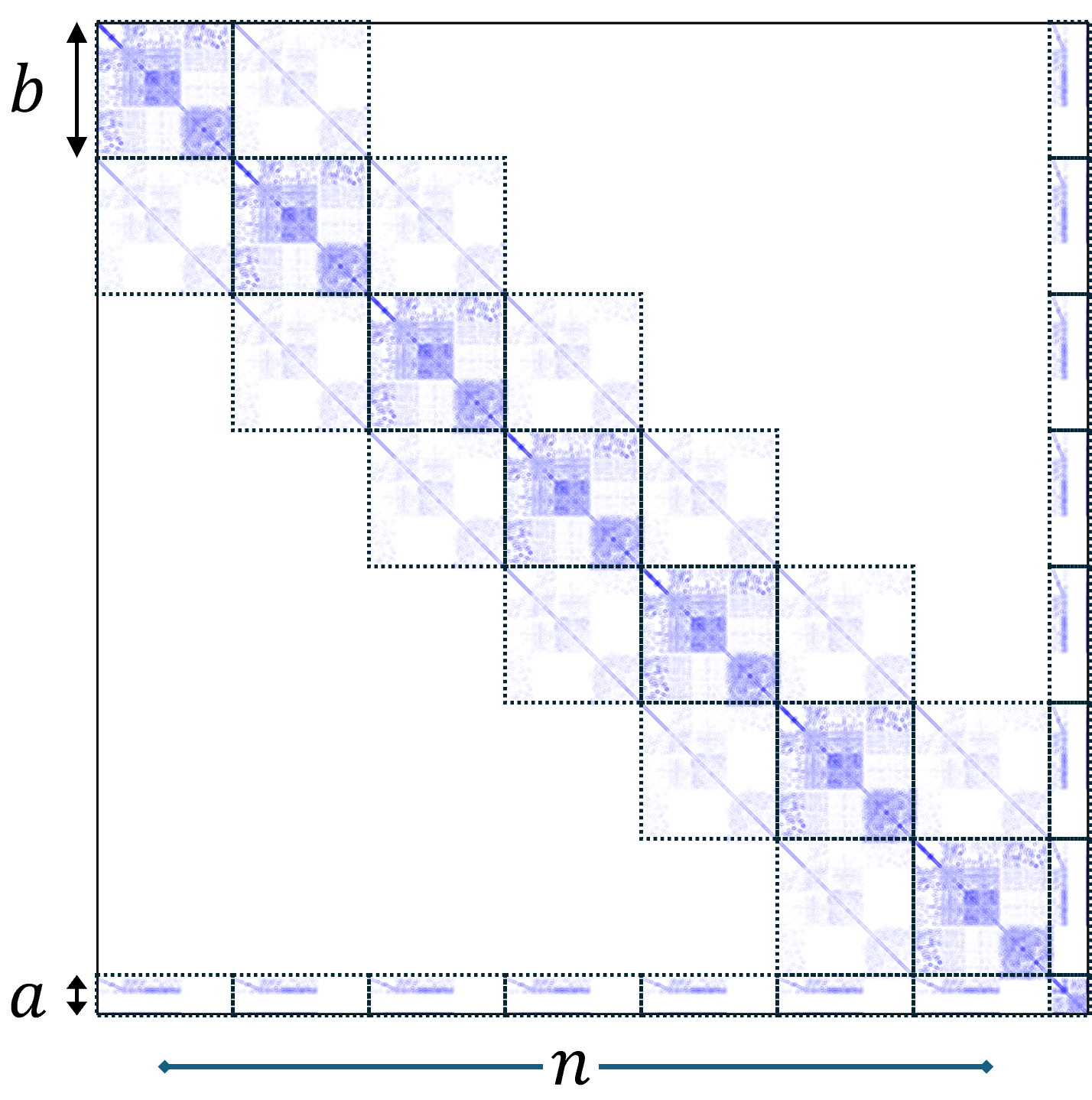}
    \end{subfigure}
    \caption{Symmetric, positive-definite, block-tridiagonal arrowhead (BTA) matrix resulting from statistical modeling applied to temperature prediction. The data was discretized on a 7-day time grid.
    % Only the lower part is shown.
    The matrix is described by the number of \textit{main diagonal blocks} $n$, their size $b$, and the arrow tip block size $a$.}
    % The total size of the matrix is $N = nb+a$.}
    \label{fig:spy_bta}
\end{figure}

%   # State-of-the-art (0.5
%\subsection{Related work}
% Differents methods
There are four types of selected-inversion algorithms (SIA):
% \begin{itemize}[leftmargin=*]
\subsubsection{Decomposition-based SIA} rely on an appropriate factorization of the initial matrix. 
They can be applied to both unstructured and block-structured sparse matrices.
In the literature, they are referred to as \emph{Takahashi} SIA.
Our methods belong to this category.
\subsubsection{Schur complement-based SIA} take advantage of the iterative (or recursive) calculation of the Schur complement of the matrix to be inverted.
They have attracted wide attention from the device modeling community due to their simplicity and capability to embed the solution of an equation of form $A X A^{\dagger} = B$ alongside the selected inversion of $A$~\cite{rgf_1, rgf_2}.
\subsubsection{Sherman-Morrison-Woodbury (SMW)-based SIA} use a divide-and-conquer approach to update the independent partitions of the matrix of interest with the SMW formula.
They are at the core of the \emph{SPIKE} algorithm and its derivatives~\cite{pdiv_1,pdiv_2,pdiv_3,pdiv_4,pdiv_gpu,splitsolve}.
SMW-based SIA are very similar to the Schur-complement ones.
\subsubsection{Block cyclic reduction (BCR)-based SIA} combine a BCR and production phase to compute the SI of a BT matrix in a divide-and-conquer fashion.
They have been extensively studied to solve tridiagonal systems of equations~\cite{bcr_1, bcr_2} and their block variants.
Because the BCR phase presents a higher complexity than typical decomposition algorithms, it limits the practical relevance of this approach.
% \end{itemize}

% Implementations
Current implementations of selected-inversion algorithms are either restricted to shared- and distributed-memory CPU architectures \cite{mumps_1,mumps_2,mumps_3,lin_selinv, lin2009fast,petersen_hybrid_2009,rgf_1, rgf_gpu} or to single GPUs \cite{gaedkeIntegrated2024}, as summarized in Table~\ref{tab:litterature_position}. As such, they do not allow for the handling of large matrices and hinder the investigation of realistic physical systems. Also, the algorithms targeting unstructured matrices generally perform sub-optimally when dealing with BT or BTA matrices. Hence, dedicated approaches have been developed for these matrix types. 
%Of great interest. the optimized shared-memory implementation of selected-inversion for BT and BTA sparse matrices (see Table~\ref{tab:litterature_position}) scale linearly with the number of diagonal blocks and provide excellent computational performance \cite{kuzmin}, despite their limitation to single nodes. 
% In the case of sparse matrices, the \emph{takahashi} selected-inversion algorithm is implemented in PARDISO and MUMPS as CPU-only solvers.
However, distributed-memory schemes only exist for the BT sparsity pattern, not for the BTA one \cite{petersen_hybrid_2009,pdiv_1,bcr_1}. 
Here, we go one step further and present distributed-memory, selected-inversion algorithms for positive semi-definite BTA matrices.
Our approach is based on a block-Cholesky decomposition and selected inversion of BTA matrices. It is implemented in a scalable library called \textit{Serinv}, which is adapted to CPU as well as GPU architectures. 
Our innovations are highlighted in cyan in Table~\ref{tab:litterature_position}. \textit{Serinv} bridges the remaining gap between the BTA structured sparsity pattern and modern, GPU-accelerated, distributed-memory algorithms.
We provide a theoretical analysis of our method and experimental results on CPU and GPU, on shared- and distributed- memory systems.
%the ALPS supercomputer at CSCS~\cite{alps_cscs,nccl_alps}.
Our new approach surpasses state-of-art, where we exhibit up to 2.6x (resp. 71.4x) speedup on CPU (resp. GPU)  over PARDISO and 14.0x (resp. 380.9x) speedup over MUMPS when scaling to 16 processes.
We achieve 32.3\% strong and 47.2\% weak scaling efficiency when going from 1 to 16 GPUs.
% This allows us to efficiently handle previously unfeasible, more realistic, problem sizes.

\begin{table}[t]
\scriptsize
\centering
\rowcolors{2}{gray!15}{white}
% \begin{tabular}{p{2.0cm}p{5.8cm}}
\begin{tabular}{p{1.78cm}p{6.25cm}}
\toprule
\textbf{Name} & \textbf{Description}\\
\midrule
$n$ & Number of square blocks in a BTA matrix's main block diagonal, excluding the arrow tip block. \\
$b$ & Size of the square blocks in a BTA matrix's main, upper, and lower block diagonals, excluding the arrowhead blocks.\\
$a$ & Size of the arrow tip block. \\
$N$ & Total size of the BTA matrix, equal to $nb + a$.\\
$P$ & Number of parallel processes and matrix partitions.\\
\midrule
\multirow{2}{*}{Block-Sequential} & Refers to algorithms and routines expressed in matrix-block operations with \emph{sequential} dependencies among them.\\
\multirow{2}{*}{Forward Pass} & Block-sequential for-loop operating from the top-left blocks of (the partition of) a matrix towards the bottom-right ones.\\
\multirow{2}{*}{Backward Pass} & Block-sequential for-loop operating from the bottom-right blocks of (the partition of) a matrix towards the top-left ones.\\
\multirow{2}{*}{True Inverse} & Refers to the blocks of a matrix's selected inverse, distinct from any transient or temporary results.\\
\midrule
PO & Positive-definite matrix.\\
BTA & Block tridiagonal arrowhead matrix.\\
F & Factorization.\\
I & Inversion.\\
SIA & Selected-inversion algorithm.\\
P & Parallel algorithm.\\
\multirow{2}{*}{PARTIAL} & Refers to an algorithm that only performs a restricted set of its normal operations.\\
\multirow{2}{*}{PERMUTED} & Refers to an algorithm acting on middle (permuted) partitions of a BTA matrix.\\
 \bottomrule
\end{tabular}
\vspace{1pt}
\caption{Symbols and terms used in this work.}
\label{tab:notation}
\end{table}

% Conclusion of intro
The main contributions of this work are the following:
\begin{itemize}[leftmargin=*]
    \item Derivation of distributed Cholesky decomposition and SIA for positive semi-definite BTA matrices;
    \item Distributed-memory implementation of these algorithms for BTA matrices on CPU and GPU;
    \item Theoretical complexity analysis of the proposed methods;
    \item Comparison with the state-of-the-art sparse solvers PARDISO and MUMPS on real-world datasets;
    \item Demonstration of strong and weak scaling of the selected-inversion algorithms.% up to 32 GPUs;
    % with an efficiency of XX\% and YY\% when going from 2 to 32 GPUs, respectively;
    % \item Application of selected inversion on large-scale, spatio-temporal temperature prediction. 
\end{itemize}
The paper is organized as follows: In Section~\ref{sec:background}, we introduce the required mathematical and algorithmic background. Our new algorithms are presented in Section~\ref{sec:methods} before conducting a theoretical analysis in Section~\ref{sec:theoretical_analysis}. We discuss our numerical results for the sequential and distributed codes in Section~\ref{sec:results}. Finally, conclusions are drawn in Section~\ref{sec:conc}.

% %   - BTA parametrization (1 ~figure)
% \begin{figure}[t]
%     \centering
%     \begin{subfigure}[t]{0.65\columnwidth}
%         \includegraphics[,width=\textwidth]{figs/spy_bta.png}
%     \end{subfigure}
%     \caption{Symmetric, positive-definite, block-tridiagonal arrowhead (BTA) sparse matrix resulting from statistical modeling applied to temperature prediction. The data was discretized on a time grid of 7 days.
%     % Only the lower part is shown.
%     The matrix is described by the number of \textit{diagonal blocks} $n$=7, their size $b$, and the arrowhead block size $a$. The total size of the matrix is $N = nb+a$.}
%     \label{fig:spy_bta}
% \end{figure}

\section{Background}
% \& Related Work}
\label{sec:background} % 2 columns

This section presents the notation and terminology used throughout the paper.
A summary is given in Table~\ref{tab:notation}.
Subsequently, we describe the current state-of-the-art methods for the selected inversion of BTA matrices.

\begin{algorithm}[t]
\small
\caption{\emph{POBTAF:} Block-sequential block-Cholesky factorization of a BTA matrix.}\label{alg:pobtaf}
\begin{algorithmic}[1]
    \Require $A$: BTA matrix. %$\Re[x^{\dagger}Ax]>0, x \in \mathbb{C}^{n}\geq 0$
    % \Require $n > 0$
    \Ensure $L$: Lower-triangular factor.
    \For{$i = 0; i < n-1; i{++}$}
        \State $L_{i, i} \gets \text{POTRF}(A_{i, i})$
        \State $L_{i+1, i} \gets \text{TRSM}(L_{i, i}, A_{i+1, i})$
        \State $L_{n, i} \gets \text{TRSM}(L_{i, i}, A_{n, i})$
        \State $A_{i+1, i+1} \gets A_{i+1, i+1} - L_{i+1, i}\times L_{i+1, i}^{\dagger}$
        \State $A_{n, i+1} \gets A_{n, i+1} - L_{n, i}\times L_{i+1, i}^{\dagger}$
        \State $A_{n, n} \gets A_{n, n} - L_{n, i}\times L_{n, i}^{\dagger}$
    \EndFor
    \State $L_{n-1, n-1} \gets \text{POTRF}(A_{n-1, n-1})$
    \State $L_{n, n-1} \gets \text{TRSM}(L_{n-1, n-1}, A_{n, n-1})$
    \State $A_{n, n} \gets A_{n, n} - L_{n, n-1}\times L_{n, n-1}^{\dagger}$
    \State $L_{n, n} \gets \text{POTRF}(A_{n, n})$
\end{algorithmic}
\end{algorithm}

\subsection{Notation and terminology}
\label{sec:notation}
In this work, we focus on BTA matrices with a sparsity pattern similar to Fig.~\ref{fig:spy_bta}.
A BTA matrix comprises three block diagonals (main, upper, and lower) and an arrowhead consisting of the last (by convention) block row and block column.
We use the term \emph{arrow tip} for the block at the intersection of the arrowhead's block row and column.
Elements within the arrowhead usually embed properties shared by (or connecting) all interacting elements in the model.
The BTA matrix's main block diagonal consists of $n$ square blocks of size $b$, while the arrow tip has size $a$.
The rest of the arrowhead blocks are rectangular with size $a\times b$ or $b\times a$.
The total size of the matrix $N$ is equal to $nb+a$.
We note that BTA matrices can be seen as a generalization of the BT sparsity pattern.
Indeed, a BT matrix is BTA with $a = 0$.

% In this work, we focus on positive-definite, block-tridiagonal arrowhead matrices, as shown in Fig.~\ref{fig:spy_bta}. 
% Such matrices can be seen as a generalization of the BT sparsity pattern by the addition of an arrowhead shape to the banded structure.
% Elements within the arrowhead usually embed properties shared by (or connecting) all interacting elements in the model.
%
While BTA matrices are ubiquitous in many fields, we focus on those arising from spatio-temporal statistical models for temperature prediction.
Fig.~\ref{fig:spy_bta} depicts a precision matrix generated using a stochastic partial differential equations approach~\cite{lindgren2011explicit} from statistical modeling.
The main diagonal blocks of the matrix correspond to the spatial discretization of the simulated domains using a finite-element method at different time steps, which are coupled through the upper and lower diagonal blocks.
% Each diagonal block corresponds to the later spatial discretization at different time steps.
% From a modeling perspective, theses diagonal components encode \emph{local effects} of the model.
While the block tridiagonal part of the matrix, relates to local phenomena, the arrowhead component accounts for global effects in the model.
% Unlike the diagonal blocks of the matrix, the arrowhead components represent global properties of the model.
% For example, and in the case of a temperature prediction model, the arrowhead components could encode elevation at different points of the spatial discretization, acting globally on the temperature trend at this altitude. 
For example, in the case of a temperature prediction model, the arrowhead component could encode information on the elevation of different space-time variables, which is assumed to affect the temperature independently of the exact time and space location. 

Since the matrices describing such problems are symmetric positive-definite, we focus on selected inversion through lower-triangular block-Cholesky factorization~\cite{block-cholesky-louter-nool}.
However, all algorithms presented here have a straightforward extension to general BTA matrices using block-LU decomposition, provided they exhibit certain properties, such as block-diagonal dominance~\cite{block-lu-demmel}.
This is a common case in materials sciences. 
% The BTA matrix's main block-diagonal consists of $n$ \textit{diagonal blocks} of size $b$, and an \textit{arrowhead block} of size $a$.
% % In the following sections, we define $n$ as being the number of diagonal blocks, $b$ the diagonal block size, and $a$ the arrowhead block size of the BTA matrix.
% The total size of the matrix $N$ is equal to $n*b+a$.

\begin{algorithm}[t]
\small
\caption{\emph{POBTASI:} Block-sequential selected inversion of a BTA matrix, given its block-Cholesky decomposition.}\label{alg:pobtasi}
\begin{algorithmic}[1]
    \Require $L$: Lower-triangular factor.
    \Ensure $X$: Selected inverse of $A$.
    \State $X_{n,n} \gets L_{n,n}^{-\dagger}\times L_{n,n}^{-1}$
    \State $U_{n,n-1} \gets -X_{n,n}\times L_{n,n-1}$
    \State $X_{n,n-1} \gets \text{TRSM}(L_{n-1,n-1}, U_{n,n-1})$
    \State $U_{n-1,n-1} \gets L_{n-1,n-1}^{-\dagger}-X_{n,n-1}^{\dagger}\times L_{n,n-1}$
    \State $X_{n-1,n-1} \gets \text{TRSM}(L_{n-1,n-1}, U_{n-1,n-1})$
    \For{$i = n-2; i >= 0; i{--}$}
        \State $U_{i+1,i} \gets -X_{i+1,i+1}\times L_{i+1,i} - X_{n,i+1}^{\dagger}\times L_{n,i}$
        \State $X_{i+1,i} \gets \text{TRSM}(L_{i,i}, U_{i+1,i})$
        \State $U_{n,i} \gets -X_{n,i+1}\times L_{i+1,i} - X_{n,n}\times L_{n,i}$
        \State $X_{n,i} \gets \text{TRSM}(L_{i,i}, U_{n,i})$
        \State $U_{i,i} \gets L_{i,i}^{-\dagger} - X_{i+1,i}^{\dagger}\times L_{i+1,i} - X_{n,i}^{\dagger}\times L_{n,i}$
        \State $X_{i,i} \gets \text{TRSM}(L_{i,i}, U_{i,i})$
    \EndFor
\end{algorithmic}
\end{algorithm}

All our methods rely on block algorithms and use BLAS and LAPACK operations.
We refer to those methods or subroutines that exhibit \emph{sequential} data dependencies among the block-level operations as \emph{block-sequential}.
These methods apply to matrices that can grow very large, rendering them amenable to asymptotic analysis.
However, the block sizes $a$ and $b$ are finite and relatively small ($\approx100-10,000$) in practical applications.
Therefore, considering the cost of distributed-memory communication in modern architectures, block-level parallelism may only be fruitfully exploited in shared memory.
The block-sequential algorithms typically consist of block-sequential for-loops called either \emph{forward} or \emph{backward} passes.
A forward pass, by convention, operates \emph{sequentially} from the top-left blocks of a matrix (or a matrix's partition, in the case of distributed-memory algorithms) towards the bottom-right ones.
A backward pass goes in the opposite direction.
The overall selected-inversion process produces intermediate results, and certain matrix blocks may be updated several times, especially in the case of in-place and distributed-memory implementations.
To distinguish between those transient results and the final output, we describe blocks of the inverse as \emph{true inverse}.

\begin{figure*}[t]
    \centering
    \includegraphics[width=\textwidth]{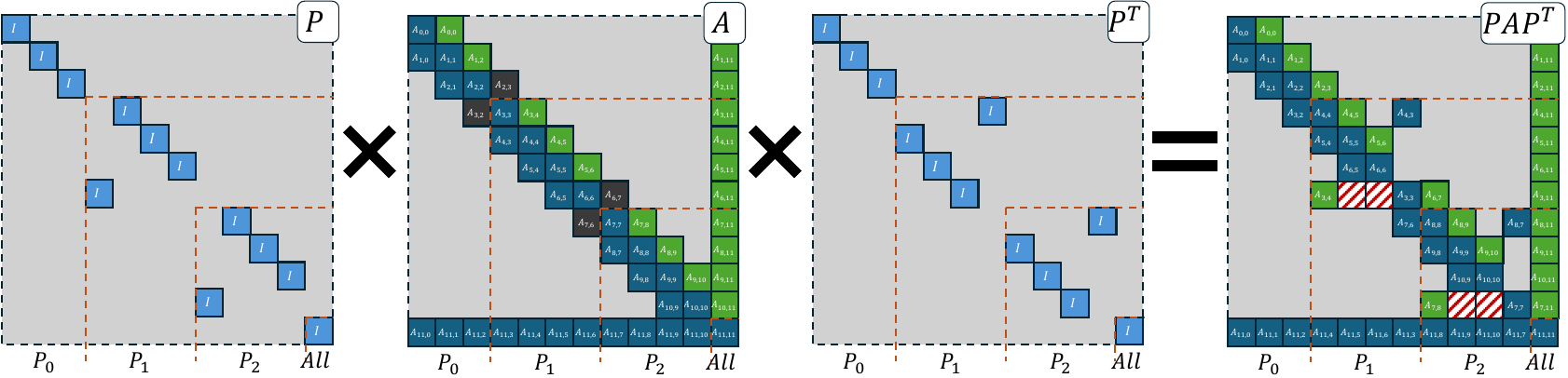}
    \caption{Permutation scheme applied on a symmetric, positive-definite, matrix in order to perform its parallel factorization and selected-inversion. The matrix is distributed amongst three processes and permuted accordingly. The permutation-induced fill-in during the decomposition is shown in red hatches in the permuted matrix.}
    \label{fig:permutation_pobta}
\end{figure*}

We follow a BLAS- and LAPACK-like naming scheme for the methods' names.
% For example, one can invert a dense positive-definite matrix using the LAPACK routines \emph{POTRF} and \emph{POTRI}.
In the LAPACK routines \emph{POTRF} and \emph{POTRI}, ``PO'' stands for positive-definite matrix, ``TR'' for triangular matrix, ``F'' for factorization, and ``I'' for inversion.
Thus, we call the Cholesky factorization of a BTA matrix \emph{POBTAF} and its selected inversion \emph{POBTASI}.
We use the ``SI'' string to explicitly distinguish selected inversion from regular full inversion.
In the case of distributed-memory algorithms, we prepend ``P'' to the routine names, similarly to PBLAS and ScaLAPACK.
We also use $P$ to define the total number of parallel processes and partitions.
Finally, we use the ``PERMUTED'' string to refer to an algorithm operating on a BTA matrix's permutation or slice.

\subsection{Block-sequential selected inversion of BTA matrices}
\label{sec:background-sequential-methods}

%   - BTA block-Cholesky algorithm (sequential) (0.5 ~algs)
The Cholesky decomposition of a symmetric, positive-definite, matrix $A$ is a lower (upper) triangular matrix $L$ ($U$) such that $A=LL^{T}$ ($A=U^{T}U$).
%
% Many standard libraries offer highly efficient implementations of the Cholesky decomposition for dense (\emph{LAPACK}~\cite{}, \emph{scaLAPACK}~\cite{}, \emph{cuBLAS}~\cite{}, \emph{rocBLAS}~\cite{}, or \emph{MAGMA}~\cite{}) and sparse matrices (PARDISO~\cite{}, MUMPS~\cite{}, and \emph{SuiteSparse}~\cite{}). 
%
Algorithm~\ref{alg:pobtaf} presents POBTAF, the block-sequential method for the block-Cholesky factorization of BTA matrices.
It is implemented in the INLA$_{\text{BTA}}$ subcomponent of the INLA$_{\text{DIST}}$ package~\cite{gaedkeIntegrated2024}.
% We name this algorithm \emph{pobtaf}, following a similar naming scheme as in LAPACK.
% Alg.~\ref{alg:pobtaf} presents a block-Cholesky algorithm tailored to BTA sparse matrices.
This algorithm executes a forward pass and, in each iteration, computes the lower-triangular Cholesky decomposition $L_{i, i}$ of the corresponding main diagonal block using LAPACK's POTRF routine.
% Sequentially and for each diagonal block of the matrix, the algorithm first computes its Cholesky decomposition using the dense LAPACK function \emph{potrf}.
The computed main diagonal factor $L_{i, i}$ is then used to determine the lower diagonal $L_{i+1, i}$ and arrowhead $L_{n, i}$ factors, which lie in the same block column.
After evaluating those factors, POBTAF performs a forward update of the next main diagonal and arrowhead blocks, $A_{i+1, i+1}$ and $A_{n, i+1}$.
It also updates the arrow tip $A_{n, n}$.
% The block-Cholesky decomposition algorithm for BTA sparse matrices is denoted by $pobtaf$.

%   - BTA block-selected-inversion algorithm (sequential) (0.5 ~algs)
% The standard procedure to perform selected inversion on general sparse matrices has been derived by Takahashi et al.~\cite{takahashi_sellinv}.
% Their selected inversion method involves computing each row (respectively column) of $A^{-1}$ until the index of the first non-zero element of $A$.
% It can be viewed as a range-restricted version of the dense inversion algorithm, as it performs the inversion until the last non-zero elements of the column (respectively row).
%
Algorithm~\ref{alg:pobtasi} introduces POBTASI, a block-sequential method for the block-selected inversion of BTA matrices, given their block-Cholesky factorization. It is also available in the INLA$_{\text{BTA}}$ package~\cite{gaedkeIntegrated2024}.
POBTASI computes $A^{-1}$'s true inverse blocks with the exact same coordinates as the non-zero blocks of $A$. 
% Alg.~\ref{alg:pobtasi} presents a block-selected-inversion algorithm tailored to BTA sparse matrices.
The algorithm performs a backward pass, iterating over the block rows of $A$, while computing the true inverse lower diagonal and arrowhead blocks, $X_{i+1,i}$ and $X_{n,i}$.
Both blocks are then used to produce the next main diagonal block of the selected inverse, $X_{i,i}$.
% Note that the presented \emph{pobtasi} algorithm computes the selected inverse, matching the structured sparsity of the initial system matrix $A$.
POBTASI is derived from the Takahashi selected-inversion method for general sparse matrices~\cite{takahashi_sellinv}, and, if desired, can be amended to compute more entries of the inverse $A^{-1}$ than the number of non-zeros in $A$.

%   - Related work on distributed selected-inversion (0.5 ~Schur equations, ~figure)
%       - BT Schur complement 
%       - BT distributed Schur complement + related work

% \begin{figure}[t]
%     \centering
%     \begin{subfigure}[t]{0.60\columnwidth}
%         \includegraphics[width=\textwidth]{figs/schur_complement.png}
%     \end{subfigure}
%     \caption{Illustration of the iteration $i$ of the Schur complement algorithm applied to a matrix with a BTA structured sparsity pattern. $M_{i,i}$ refers to the Schur complement of the slice $A_{0:i, 0:i}$ of the initial matrix, while $M_{i+1,i+1}$ corresponds to the forward update of the slice $A_{i+1:n-1, i+1:n-1}$ calculated at the next iteration of the method. The update is computed as follows $M_{i+1,i+1} \gets M_{i+1,i+1} - M_{i+1, i} M_{i, i}^{-1} M_{i, i+1}$.}
%     \label{fig:schur_complement}
% \end{figure}

\section{Methods}
\label{sec:methods} % 4 columns (~fig permutation, ~fig pipeline)
To break the data dependencies of the block-sequential algorithms presented in Section~\ref{sec:background-sequential-methods}, we introduce a permutation scheme extending the work of Petersen et al.~\cite{petersen_hybrid_2009}.
Applying this permutation to the input matrix allows us to reorder the block operations, exposing parallel sections in the forward and backward passes.
Overall, the algorithm we propose splits the input matrix $A$ into $P$ partitions and comprises three phases: (1) \emph{PPOBTAF} computes \emph{partial} block-Cholesky factorizations for all partitions in parallel.
(2) \emph{POBTARSSI} gathers the partial factorizations' blocks lying in the boundaries among the partitions to construct a smaller BTA matrix called \emph{reduced system} and labeled $A_r$. The method then computes the reduced system's selected inverse by sequentially applying the \emph{POBTAF} and \emph{POBTASI} methods. The selected inverse of the reduced-system, $X_r$, consists of the true inverse boundary blocks of each partition.
(3) \emph{PPOBTASI} uses the latter blocks to execute the backward pass of Algorithm~\ref{alg:pobtasi} for all partitions in parallel, which delivers selected entries of $A^{-1}$.

% The algorithm we propose comprises three phases and can be defined as a divide-and-conquer approach.
% The first and last phases are embarrassingly parallel, while the middle step consists of the assembly through communication and inversion of a reduced system with the $d\_pobtarssi$) routine.
% To break the complete sequential dependencies\footnote{During the Cholesky factorization, it is possible to compute the lower diagonal block $A_{i+1,i}$ and arrowhead block $A_{n,i}$ in parallel. However, this strategy does not present any advantage as the computations are memory-bound and streaming both blocks to the GPU for parallel processing does not provide any speed-up} of the block algorithms presented in Section~\ref{sec:background-sequential-methods} and expose parallelism in the block Cholesky-factorization and selected inversion procedures, new computational schemes must be introduced, namely permutation and re-ordering of the matrix.
% We present a permutation scheme, extending the work of Petersen et al.~\cite{petersen_hybrid_2009}, that enables embarrassingly parallel operations during the factorization and selected inversion of BTA matrices.

%   - Control flow and processes slicing (0.25)
\subsection{Partitioning and permutation schemes}
\label{sec:partitioning-pemutation}
We partition the BTA matrix into $P$ disjoint arrow shapes, each consisting (approximately) of $n_p = n / P$ block rows and columns.
% A process $p$ is assigned the slices $[p*n_{p}:(p+1)*n_{p}]$ of the main diagonal, lower diagonal, and arrowhead blocks.
The coordinates of the blocks belonging to a partition $p$ are $A_{i, i}$ (main block diagonal), $A_{i+1,i}$ (lower block diagonal), and $A_{n, i}$ (arrowhead) for $i \in \{pn_p .. (p+1)n_p-1\}$.
We distinguish between the first (top) partition $p_0$, which owns block $A_{0, 0}$, and the rest, \emph{middle} partitions.
All partitions share the arrow tip block $A_{n, n}$ and can then be re-interpreted as a BTA matrix with $n=n_p$.
% An example BTA matrix, with $n=11$, split into three partitions is shown in Fig.~\ref{fig:pipeline}.a).
An example BTA matrix, with $n=11$, split into three partitions is shown, alongside its permutation matrix, in Fig.~\ref{fig:permutation_pobta}.

%
% The first process operates on its partition using the sequential $pobtaf$ algorithm, while the other processes (referred to as \emph{middle processes}) first permute their local partition.

%   - Permutation (~fig permutation) (0.5)
% We distinguish between the first (top) partition $p_0$, which owns block $A_{0, 0}$, and the rest, \emph{middle} partitions.
% Our permutation scheme applies only to the latter.
% Figure~\ref{fig:permutation_pobta}.a) shows a middle partition with $n_p=11$, re-interpreted as an individual BTA matrix $A$.
% The corresponding permutation matrix $P$ is depicted in Fig.~\ref{fig:permutation_pobta}.b), while Fig.~\ref{fig:permutation_pobta}.c) presents the resulting permuted matrix $PAP^T$.
% Figure~\ref{fig:permutation_pobta} shows the permutation matrix applied to the partitioned BTA matrix.
% The corresponding permutation matrix $P$ is depicted in Fig.~\ref{fig:permutation_pobta}.b), while Fig.~\ref{fig:permutation_pobta}.c) presents the resulting permuted matrix $PAP^T$.
% The permutation applied to the middle partitions is presented in Fig.~\ref{fig:permutation_pobta}.
% It connects the middle partitions upward (to the previous partition) and downward (to the following partition) at the end of the decomposition.
The matrix $P$ can be defined as a \emph{shifting} permutation, re-organizing the blocks inside the sparsity structure.
In practice, applying this permutation matrix to a BTA partition shifts its blocks up and left along the diagonal.
The first diagonal block belonging to this partition is shifted around to the last position.
For example, in Fig.~\ref{fig:permutation_pobta}, the first diagonal block of partition 1 is $A_{3,3}$.
After applying the permutation, the first diagonal block is $A_{4,4}$, while $A_{3,3}$ is shifted around to the last position.
This permutation is necessary to factorize the middle partitions toward their two neighboring partitions, above and below, allowing for further completion of the partial factorization during the \emph{POBTARSSI} step.
The blocks connecting the partitions, represented in dark gray color in Fig.~\ref{fig:permutation_pobta}, remain untouched during the permutation.
These blocks are latter going to be used to connect the partitions together when assembling the reduced system at the end of the parallel factorization phase.
The middle partitions' permutation induces fill-in during the partial block-Cholesky factorization, annotated with red hatches in the permuted matrix in Fig.~\ref{fig:permutation_pobta}.
% The fill-in induced in the Cholesky factorization because of the permutation is shown in red hatches.
Due to this fill-in, the PPOBTAF and PPOBTASI methods execute more block operations for the middle partitions than for the first one.
Load balancing issues are further discussed in Section~\ref{sec:theoretical_analysis}.
We note that the permutation is never materialized but remains implicit, i.e., the blocks of $A$ and $L$ are not reordered in memory.

\begin{algorithm}[t]
\small
\caption{\emph{PPOBTAF:} Parallel block-Cholesky factorization of a BTA matrix.}
\label{alg:ppobtaf}
\begin{algorithmic}[1]
    \Require $A_{p_i}, i \in \{0, ..., P-1\}$: $P$ partitions of $A$.
    \Ensure $L_{p_i}, i \in \{0, ..., P-1\}$: $P$ partitions of $L$.
    \Ensure $B_{p_i}, i \in \{1, ..., P-1\}$: $P-1$ fill-in partitions.
    % \ForAll{$i=0; i < P; i{++}$}
    \ForAll{$i \in \{0, ..., P-1\}$}
        \If{$i == 0$}
            \State $(L_{p_0}, U_{p_0}) \gets \text{PARTIAL\_POBTAF}(A_{p_0})$
        \Else
            \State $(L_{p_i}, B_{p_i}, U_{p_i}) \gets\text{PERMUTED\_POBTAF}(A_{p_i})$
        \EndIf
    \EndFor
    \State $L_{n,n} \gets A_{n,n} + \sum_{i = 0}^{P-1} U_{p_i}$
\end{algorithmic}
\end{algorithm}
\begin{algorithm}[t]
\small
\caption{\emph{PERMUTED\_POBTAF:} Block-sequential partial block-Cholesky factorization of a middle BTA partition. Differences compared to POBTAF are highlighted in red.}
\label{alg:permuted_pobtaf}
\begin{algorithmic}[1]
    \Require $A$: Middle partition with $n_p$ main diagonal blocks.
    \Ensure $L$: Middle lower-triangular factor partition.
    \Ensure $B$: Permutation-induced fill-in.
    \Ensure $U_{tip}$: Partial arrow tip update.
    \State \textcolor{red}{$B_1 = A_{1,0}^{\dagger}$, $U_{tip} = 0$}
    \For{$i = 1; i < n_p-1; i{++}$}
        \State $L_{i, i} \gets \text{POTRF}(A_{i, i})$
        \State $L_{i+1, i} \gets \text{TRSM}(L_{i, i}, A_{i+1, i})$
        \State $L_{n_p, i} \gets \text{TRSM}(L_{i, i}, A_{n_p, i})$
        \State \textcolor{red}{$B_{i} \gets \text{TRSM}(L_{i, i}, B_{i})$}
        \State $A_{i+1, i+1} \gets A_{i+1, i+1} - L_{i+1, i}\times L_{i+1, i}^{\dagger}$
        \State $A_{n_p, i+1} \gets A_{n_p, i+1} - L_{n_p, i}\times L_{i+1, i}^{\dagger}$
        \State \textcolor{red}{$U_{tip} \gets U_{tip} - L_{n_p, i}\times L_{n_p, i}^{\dagger}$}
        \State \textcolor{red}{$A_{0, 0} \gets A_{0, 0} - B_{i}\times B_{i}^{\dagger}$}
        \State \textcolor{red}{$B_{i+1} \gets - B_{i}\times L_{i+1, i}^{\dagger}$}
        \State \textcolor{red}{$A_{n_p, 0} \gets A_{n_p, 0} - L_{n_p, i}\times B_{i}^{\dagger}$}
    \EndFor
    % \State \textcolor{red}{$A_{n_p,n_p} \gets A_{n_p,n_p} + \sum_{i = 0}^{p} U_{i}$}
\end{algorithmic}
\end{algorithm}

%   - Algorithm listing (~permuted Cholesky) (0.75)
% \subsection{Parallel partial block-Cholesky factorization of BTA matrices}
\subsection{Parallel partial BTA block-Cholesky factorization}
\label{sec:ppobtaf}

The first step in our parallel selected-inversion algorithm is to produce a partial block-Cholesky factorization of the initial BTA matrix.
To that end, we split the input into $P$ partitions using the scheme presented in Section~\ref{sec:partitioning-pemutation}, one top partition, and $P-1$ middle partitions.
The factorizations' computations are independent and, therefore, we can employ $P$ processes in parallel, as shown in Algorithm~\ref{alg:ppobtaf}.
We use different block-sequential algorithms for the top (\emph{PARTIAL\_POBTAF}) and middle partitions  (\emph{PERMUTED\_POBTAF}).
Both routines are variations of POBTAF, specifically lines 1--8.
The main difference is that, instead of updating the arrow tip block at each step, the update is accumulated in a buffer $U_{tip}$.
At the end of the decomposition, the buffers from all partitions are added together.
In addition to the normal operations arising in the partial factorization, the permuted factorization needs to compute extra blocks during the factorization.
After applying the permutation scheme, the lower (resp. upper) blocks connecting the first diagonal block of the partition to the next one are shifted at the extremity of the partition.
For example, in Fig.~\ref{fig:permutation_pobta} and after application of the permutation matrix, the block $A_{4,3}$ (resp. $A_{3,4}$) that was initially within the BTA pattern is shifted outside of it, leading to the fill-in hatched in red during the factorization phase.
These additional computations are performed in Algorithm~\ref{alg:permuted_pobtaf} and Algorithm~\ref{alg:_permuted_pobtasi} using the buffer $B_{i}$.
The differences compared to POBTAF are highlighted in red.
% At the end of the decomposition, all buffers from all partitions are added together; the resulting update is then applied to the tip of the arrow. The fill-in induced by the permutation results in the computation of extra $L$ factors, noted in Algorithm~\ref{alg:permuted_pobtaf} as $B$.
% The $permuted\_pobtaf$ algorithm performs the decomposition starting from the second block to the penultimate block of the partition.
% This allows for the later creation of a reduced system using the updated (but not factorized) first and last blocks of the middle partitions.
% In the following listing, the $trsm$ operations are right-sided (i.e., we solve the linear system from the right side).

\begin{algorithm}[t]
\small
\caption{\emph{PPOBTASI:} Parallel selected inversion of a BTA matrix, given its block-Cholesky factorization.}
\label{alg:ppobtasi}
\begin{algorithmic}[1]
    \Require $L_{p_i}, i \in \{0, ..., P-1\}$: $P$ partitions of $L$.
    \Require $B_{p_i}, i \in \{1, ..., P-1\}$: $P-1$ fill-in partitions.
    \Ensure $X_{p_i}, i \in \{0, ..., P-1\}$: $P$ partitions of $X$.
    % \ForAll{$i=0; i < P; i{++}$}
    \ForAll{$i \in \{0, ..., P-1\}$}
        \If{$i == 0$}
            \State $X_{p_0} \gets \text{PARTIAL\_POBTASI}(L_{p_0})$
        \Else
            \State $X_{p_i} \gets\text{PERMUTED\_POBTASI}(L_{p_i}, B_{p_i})$
        \EndIf
    \EndFor
\end{algorithmic}
\end{algorithm}
%   - Algorithm listing (~permuted SellInv) (0.75)
\begin{algorithm}[t]
\small
\caption{\emph{PERMUTED\_POBTASI:} Block-sequential selected inversion of a middle BTA partition, given its block-Cholesky decomposition. Differences compared to POBTASI are highlighted in red.}
\label{alg:_permuted_pobtasi}
\begin{algorithmic}[1]
    \Require $L$: Middle lower-triangular factor partition.
    \Require $B$: Permutation-induced fill-in.
    \Ensure $X$: Middle partition with $n_p$ main diagonal blocks.
    \For{$i = n_p-2; i > 0; i{--}$}
        \State $U_{i+1,i} \gets -X_{i+1,i+1}\times L_{i+1,i} - X_{n_p,i+1}^{\dagger}\times L_{n_p,i}$
        \State \textcolor{red}{$U_{i+1,i} \gets U_{i+1,i} - B_{i+1}^{\dagger}\times B_{i}$}
        \State $X_{i+1,i} \gets \text{TRSM}(L_{i,i}, U_{i+1,i})$
        \State \textcolor{red}{$V_{i} \gets -B_{i+1}\times L_{i+1,i} -X_{0,0}\times B_{i} - X_{n_p,0}^{\dagger}\times L_{n_p,i}$}
        \State \textcolor{red}{$B_{i} \gets \text{TRSM}(L_{i,i}, V_{i})$}
        \State $U_{n_p,i} \gets -X_{n_p,i+1}\times L_{i+1,i} -X_{n_p,n_p}\times L_{n_p,i}$
        \State \textcolor{red}{$U_{n_p,i} \gets U_{n_p,i} - X_{n,0}\times L_{0,i}$}
        \State $X_{n_p,i} \gets \text{TRSM}(L_{i,i}, U_{n_p,i})$
        \State $U_{i,i} \gets L_{i,i}^{-\dagger}\times -X_{i+1,i}^{\dagger}\times L_{i+1,i} -X_{n_p,i}^{\dagger}\times L_{n,i}$
        \State \textcolor{red}{$U_{i,i} \gets U_{i,i} - B_{i}^{\dagger}\times B_{i}$}
        \State $X_{i,i} \gets \text{TRSM}(L_{i,i}, U_{i,i})$
    \EndFor
\end{algorithmic}
\end{algorithm}

\begin{figure*}[t]
    \centering
    \begin{subfigure}[t]{\textwidth}
        \includegraphics[width=\textwidth]{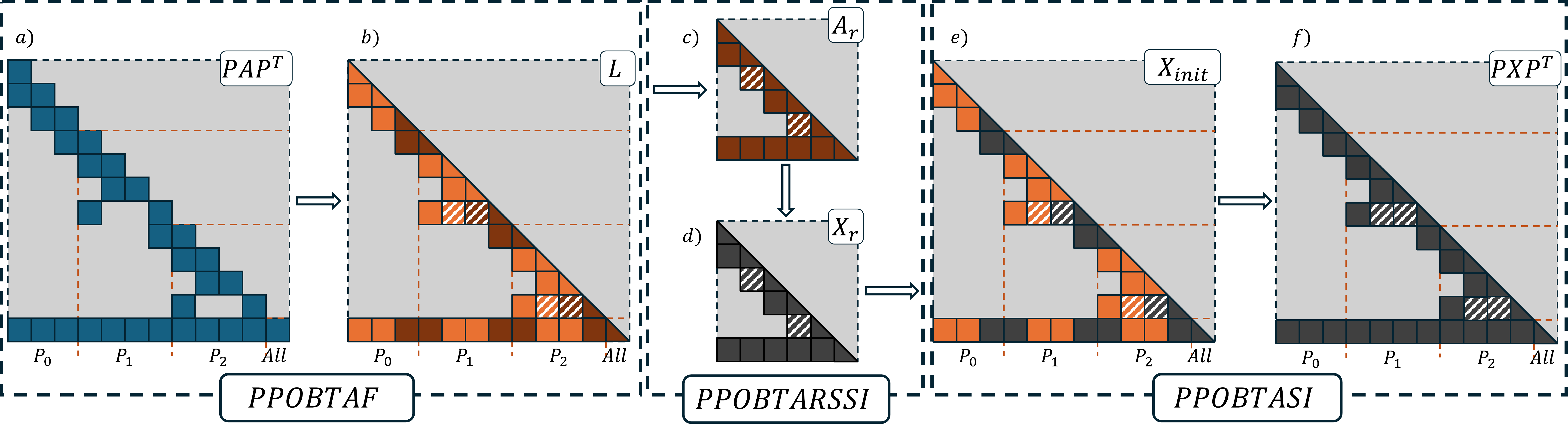}
    \end{subfigure}
    \caption{General organization of the distributed block-Cholesky factorization and selected inversion of a positive-definite BTA matrix. The method consists of three steps. a) and b) Parallel block-Cholesky factorization. c) and d) Creation of the reduced system $A_r$ and its selected inversion $X_r$. e) and f) Parallel selected inversion.}
    \label{fig:pipeline}
\end{figure*}

\subsection{Reduced system and its selected inversion}
\label{sec:pobtarssi}
%   - (~fig pipeline) (1.75)
Performing the partition-independent factorization of the matrix comes at the cost of solving an adjoint set of linear equations, i.e., a reduced system connecting the partitions. 
This reduced system, $A_{r}$, presents the same sparsity pattern as the initial matrix $A$ but with fewer blocks.
The number of blocks in the main diagonal of $A_r$ is $n_{r} = 2P-1$.
The construction of this system from the blocks lying at the boundary among the partitions (brown color) is shown in Fig.~\ref{fig:pipeline}.b) and c).
$A_r$ is solved with selected inversion, and the procedure is referred to as POBTARSSI.
This operation can be implemented either using the block-sequential selected-inversion algorithms (POBTAF and POBTASI) or by applying our parallel process (PPOBTAF, POBTARSSI, and PPOBTASI), potentially in a recursive manner.
The latter approach leads to the nested-solving method discussed in Section~\ref{sec:nested-solving}.
The reduced system's selected inverse, $X_r$, consists of the true inverse blocks lying at the boundary among the partitions (dark gray color), as shown in Fig.~\ref{fig:pipeline}.d) and e).

\subsection{Parallel selected inversion of BTA matrices}
\label{sec:ppobtasi}

Algorithm~\ref{alg:ppobtasi} describes the parallel selected inversion of a BTA matrix.
POBTARSSI's result is copied to PPOBTASI's $L$ and $B$ inputs, as illustrated in Fig.~\ref{fig:pipeline}.d) and e) in dark gray.
The selected inversion of each partition can be computed independently of the others, and we can employ $P$ processes in parallel.
Similarly to PPOBTAF, we use different block-sequential algorithms for the top and middle partitions, \emph{PARTIAL\_POBTASI} and \emph{PERMUTED\_POBTASI}, respectively.
Both routines are derived from POBTASI, specifically lines 6--13.
The middle partitions must execute more block operations due to the permutation-induced fill-in $B$, highlighted in red in Algorithm~\ref{alg:_permuted_pobtasi}.
Figure~\ref{fig:pipeline} outlines the entire procedure from the parallel factorization of the initial BTA matrix (represented as permuted) to the construction and selected inversion of the reduced system and, finally, its parallel selected inversion.

\begin{table*}[t]
\centering
\resizebox{\textwidth}{!}{%
% \begin{tabular}{@{}cccc@{}}
% \rowcolors{2}{gray!15}{white}
\begin{tabular}{llrrrrrr}
\toprule
\multirow{2}{*}{\textbf{Routine}} & \multirow{2}{*}{\textbf{Description}} & \multirow{2}{*}{\textbf{Complexity}} & \multicolumn{5}{c}{\textbf{Block-Operation Count}} \\
\cmidrule(l){4-8} & & & \textbf{POBTAF} & \textbf{POBTASI} & \textbf{PPOBTAF} & \textbf{POBTARSSI} & \textbf{PPOBTASI}\\
\midrule
\multirow{2}{*}{POTRF} & \multirowcell{2}{Cholesky\\Decomposition} & $O(a^3)$ & $1$ & $0$ & $0$   & $1$ & $0$\\
% $O(a^3)$ & $n$ & $0$ & $n/(P-2)$   & $2P-1$ & $0$
&  & \mycc$O(b^3)$ & \mycc$n$ & \mycc$0$ & \mycc$\nicefrac{n}{P}-2$ & \mycc$2P-1$ & \mycc$0$\\
\midrule
\multirow{4}{*}{GEMM} & \multirowcell{4}{General\\Matrix-Matrix\\Multiplication} & $O(a^3)$ & $0$ & $1$ & $0$ & $1$ & $0$\\
& & \mycc$O(a^2b)$ & \mycc$n$ & \mycc$n-1$ & \mycc$\nicefrac{n}{P}-2$ & \mycc$4P-3$ & \mycc$\nicefrac{n}{P}-2$\\
& & $O(ab^2)$ & $n-1$ & $n-1$ & $\nicefrac{n}{P}-2$ & $4P-4$ & $\nicefrac{n}{P}-2$\\
&  & \mycc$O(b^3)$ & \mycc$n-1$ & \mycc$n-1$ & \mycc$\nicefrac{n}{P}-2$ & \mycc$4P-4$ & \mycc$\nicefrac{n}{P}-2$\\
\midrule
\multirow{2}{*}{TRSM} & \multirowcell{2}{Triangular\\Solve} & $O(ab^2)$ & $n$ & $1$ & $\nicefrac{n}{P}-2$ & $2P$ & $0$\\
&  & \mycc$O(b^3)$ & \mycc$n-1$ & \mycc$n-1$ & \mycc$\nicefrac{n}{P}-2$ & \mycc$4P-4$ & \mycc$\nicefrac{n}{P}-2$\\
 \bottomrule
\end{tabular}
}
\vspace{5pt}
\caption{Complexity of the BLAS and LAPACK routines used to perform Cholesky decomposition (POTRF), matrix-matrix multiplication (GEMM), and triangular solve (TRSM). $a$ is the size of the arrow tip block, $b$ is the size of main and lower diagonal blocks, while the rest of the arrowhead blocks have size $a\times b$ or $b\times a$. We also list the number of times each of these functions is called in routines for selected inversion of BTA matrices: POBTAF (block-sequential Cholesky factorization), POBTASI (block-sequential selected inversion), PPOBTAF (parallel Cholesky factorization), POBTARSSI (reduced system selected inversion), and PPOBTASI (parallel selected inversion). The counts of operations match the implementation made in the \textit{Serinv} library and not necessarily the algorithmic listings.}
\label{tab:lapack}
\end{table*}

\section{Theoretical analysis}
\label{sec:theoretical_analysis} % 3 columns
%   - Lapack FLOPs (0.25 ~table)
In this section, we analyze the block-sequential and parallel algorithms derived in the previous sections.
In Table~\ref{tab:lapack}, we present the BLAS and LAPACK routines used to operate on the dense blocks of the BTA matrices.
Our algorithms rely on three routines, POTRF, GEMM, and TRSM, whose complexity is a function of the diagonal and arrow tip block sizes $b$ and $a$.
Table~\ref{tab:lapack} further lists the number of calls of those three routines per method introduced in Sections~\ref{sec:background-sequential-methods}, \ref{sec:ppobtaf}, \ref{sec:pobtarssi}, and \ref{sec:ppobtasi}.
The given block-operation count for POBTARSSI corresponds to a block-sequential implementation (POBTAF and POBTASI) executed by a single process.
We note that the parallel methods' GEMM and TSRM counts follow our implementation discussed in Section~\ref{sec:implementation}, where we have applied an optimization compared to Algorithms~\ref{alg:ppobtaf} and \ref{alg:ppobtasi}.
Concretely, these algorithms call TRSM multiple times with the same lower-triangular factor $L_{i,i}$.
Due to the performance characteristics of the TRSM and GEMM kernels executed on state-of-the-art GPU accelerators, it is beneficial to invert $L_{i, i}$ once and substitute the subsequent TRSM calls with GEMM operations.
We further discuss the \emph{practical} performance of those kernels in Section~\ref{sec:discussion}.

\subsection{Computational complexity}
\label{sec:complexity}

Both block-sequential algorithms, POBTAF and POBTASI, exhibit the same asymptotic complexity with respect to the BTA matrix parameters $a$, $b$, and $n$: $O\left(nb^3\right)$ for sufficiently large $n$ and under the assumption $a\leq b$.
The most important constant factors are 3 and 2 for POBTAF and POBTASI, respectively.
Similarly, the cost per process for the corresponding parallel algorithms is $O\left((\nicefrac{n}{P})b^3\right)$ with the same constant factors.
However, the block-sequential POBTARSSI has complexity $O(Pb^3)$ with constant factor 20, inducing a work-depth trade-off to the full parallel selected-inversion algorithm.
The total work is $O\left((n+P)b^3\right)$.
Ignoring the $b^3$ factor, the depth is $O\left(\nicefrac{n}{P} + P\right)$, limiting scaling.

\subsection{Nested solving}\label{sec:nested-solving}
Instead of executing POBTARSSI block sequentially, we can invert the reduced system using the developed parallel procedure.
We note that we have to employ a subset of the processes, as this approach always leads to another reduced system $A_{r_{ns}}$ with $2P_{ns}-1$ main diagonal blocks, where $P_{ns}$ is the number of \emph{nested-solving} processes.
If $P_{ns} = \nicefrac{P}{c}$ for a constant $c$, the algorithmic depth reduces from $O\left(\nicefrac{n}{P} + P\right)$ to $O\left(\nicefrac{n}{P} + \nicefrac{P}{c} + c\right)$.
The depth can be further decreased by using the nested-solving approach recursively.
However, every recursive call increases communication volume and, especially, latency.
Therefore, a proper theoretical evaluation of recursive nested solving must include a communication model.
In this work, we limit our exploration to calling nested solving only once, using half the processes without modeling communication.

To highlight the scaling potential in more concrete terms, we restrict the values of the parameters describing a BTA matrix.
Although $n$ can grow arbitrarily large in typical scientific problems described by BTA matrices, the same is not true for $a$ and $b$, which usually have values in the range of $1-500$ for $a$ and 100--10,000 for $b$, frequently on the lower side.
For the rest of this section, we therefore fix $a$ and $b$ to 256 and 1024, respectively.

\begin{table}[b]
\centering
\begin{tabular}{@{}lccccc@{}}
% \begin{tabular}{lccccc}
\textbf{Number of blocks}                     & $32$   & $64$   & $128$  & $256$  & $512$  \\ \midrule
\textbf{Load balancing} PPOBTAF           & $1.79$ & $1.83$ & $1.84$ & $1.85$ & $1.86$ \\
\textbf{Load balancing} PPOBTASI          & $2.43$ & $2.45$ & $2.46$ & $2.46$ & $2.46$ \\
\textbf{Ratio} PPOBTAF/PPOBTASI & $0.34$ & $0.34$ & $0.35$ & $0.35$ & $0.35$ \\ \midrule
\textbf{Ideal load balancing} ($r_{LB}$)        & $2.22$ & $2.24$ & $2.24$ & $2.25$ & $2.25$
\end{tabular}
\caption{Load balancing factor $r_{LB}$ between the partitions assigned to the \emph{top} and the \emph{middle} processes as a function of the number of blocks in the matrix. The reported values are obtained by first computing the load balancing between these partitions for the PPOBTAF and PPOBTASI routines and then by weighting the results with the number of operations performed for each function.}
\label{tab:load_balancing}
\end{table}

\subsection{Load balancing}
\label{sec:load-balancing}
As discussed in Section~\ref{sec:partitioning-pemutation}, the middle processes perform more work due to the permutation fill-in, causing workload imbalances.
To address this issue, the top partition's size can be increased.
To find the optimal ratio between the size of the top ($n_{p_0}$) and middle partitions ($n_{p_i}$), i.e., $r_{LB}$=$n_{p_0}/n_{p_i}$, we analyze in Table~\ref{tab:load_balancing} the ideal workload balance given the floating-point operation count ratio between the block-sequential and parallel algorithms.
Since the ideal load balancing factor differs for the factorization and selected-inversion algorithms, we weight it according to the ratio of operations between the PPOBTAF and PPOBTASI routines.
We report the load balancing factor for several values of $n$, ranging from $32$ to $512$, and find that it remains close to $2.25$ in all cases.

\begin{figure*}[t]
    \centering
    % \begin{subfigure}[t]{\textwidth}
        % \includegraphics[width=\textwidth]{figs/theoretical_perf_matrix.png}
    \includegraphics[width=\textwidth]{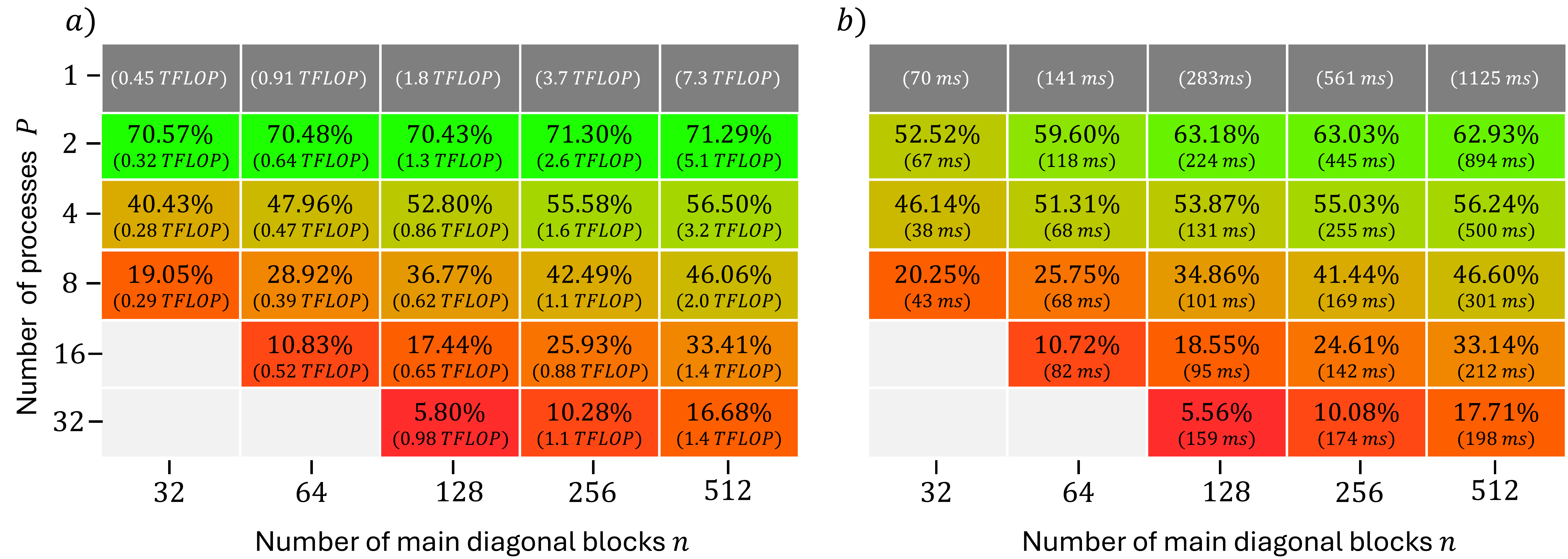}
    % \includegraphics[width=\textwidth]{figs/oldcolor_parallel_efficiency_matrix.pdf}
    % \end{subfigure}
    \caption{a) Theoretical maximum parallel efficiency of the complete selected-inversion procedure (PPOBTAF + POBTARSSI + PPOBTASI) as a function of the number of main diagonal blocks (horizontal axis) and processes (vertical axis) for BTA matrices with $b$=1024 and $a$=256. b) Experimental parallel efficiency of the complete selected-inversion procedure (PPOBTAF + POBTARSSI + PPOBTASI) using the theoretically determined ideal load balancing factor $r_{LB}$.}
    \label{fig:theoretical_perf_matrix}
\end{figure*}

\subsection{Parallel efficiency}
\label{sec:theoretical-efficiency}

Using the floating-point operation counts for the block-sequential and parallel algorithms from Table~\ref{tab:lapack} and the ideal load balancing ratios from Table~\ref{tab:load_balancing}, we present in Fig.~\ref{fig:theoretical_perf_matrix}.a) the theoretical maximum parallel efficiency of the entire parallel selected-inversion procedure (i.e., PPOBTAF + block-sequential POBTARSSI + PPOBTASI), for $P$ ranging from 1 to 32 and $n$ from 32 to 512.
For every data point, we show the workload per process in TFLOP in parentheses and compute the efficiency as its ratio to the block-sequential workload for the same $n$ value. 
The highest efficiency per process count is obtained for the maximal number of diagonal blocks $n=512$, where the ratio between the reduced system's size and each partition is the most advantageous.
For the same reason, fewer processes lead to greater parallel efficiency.

\section{Evaluation}
\label{sec:results} % 6 columns
We implement the methods described in Section~\ref{sec:methods} and evaluate their performances.
First, we measure the parallel efficiency of our methods on a synthetic dataset, giving an overview of our method's performances.
Then, we perform a weak scaling analysis for both the block-sequential and nested-solving approaches for solving the reduced system.
Finally, we showcase a comparison of our methods, against the state-of-the-art sparse direct solvers MUMPS and PARDISO.
We will refer to \emph{BTA-density} the density of non-zeros elements within the BTA sparsity pattern of a sparse matrix.

\subsection{Implementation}
\label{sec:implementation}
We implement all block-sequential and parallel algorithms in Python, using NumPy\cite{harris2020array} on CPU and CuPy~\cite{cupy}on GPU to interface with optimized BLAS/LAPACK libraries.
Our code utilizes MPI with mpi4py~\cite{mpi4py}, GPU-aware if available, but also the NVIDIA Collective Communications Library (NCCL)~\cite{nccl_1,nccl_2,nccl_3,nccl_4,nccl_alps} on machines equipped with NVLink for direct inter-GPU communication.
We employ one (MPI) process per CPU or GPU chip (socket).
As described in Section~\ref{sec:theoretical_analysis}, our implementations replace many of the TRSM operations appearing in Algorithms~\ref{alg:permuted_pobtaf} and~\ref{alg:_permuted_pobtasi} by GEMM calls, which exhibit much higher performance in the utilized BLAS libraries.
We use Reduce, Gather, and Scatter collectives (and their all-variants) to assemble the reduced system in the root process (or a subset of processes in the case of nested solving), perform its selected inversion, and scatter back the results, resulting in a gather-scatter communication strategy.

\begin{figure}[b]
\vspace{1em}
    \centering
    \begin{subfigure}[t]{0.9\columnwidth}
        \includegraphics[width=\textwidth]{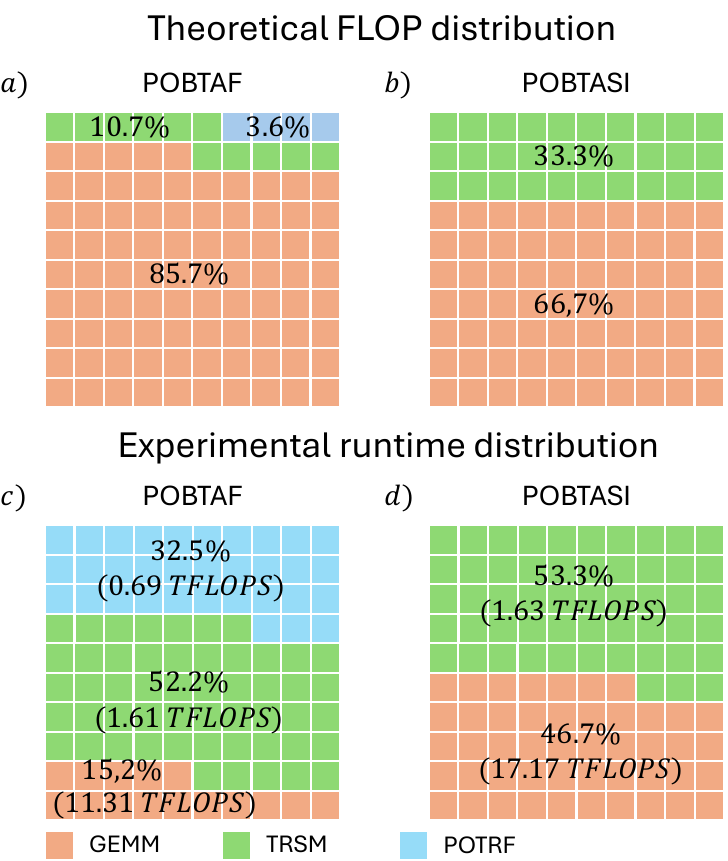}
    \end{subfigure}
    \caption{Theoretical FLOP count distributions for the GEMM, TRSM, and POTRF routines within the a) POBTAF and b) POBTASI algorithms. c) and d) report the corresponding breakdown based on actual runtime measurements for a BTA matrix with $b=1024$ and $a=256$ on an NVIDIA GH200 (GPU). We show the kernel performances in TFLOPS in parentheses.}
    \label{fig:sequential_cost_breakdown}
\end{figure}

\begin{figure*}[t]
    \centering
    % \begin{subfigure}[t]{\textwidth}
    %     \includegraphics[width=\textwidth]{figs/experimental_strong_scaling.png}
    % \end{subfigure}
    \begin{subfigure}[t]{.32\textwidth}
        \caption{}
        \includegraphics[width=\textwidth]{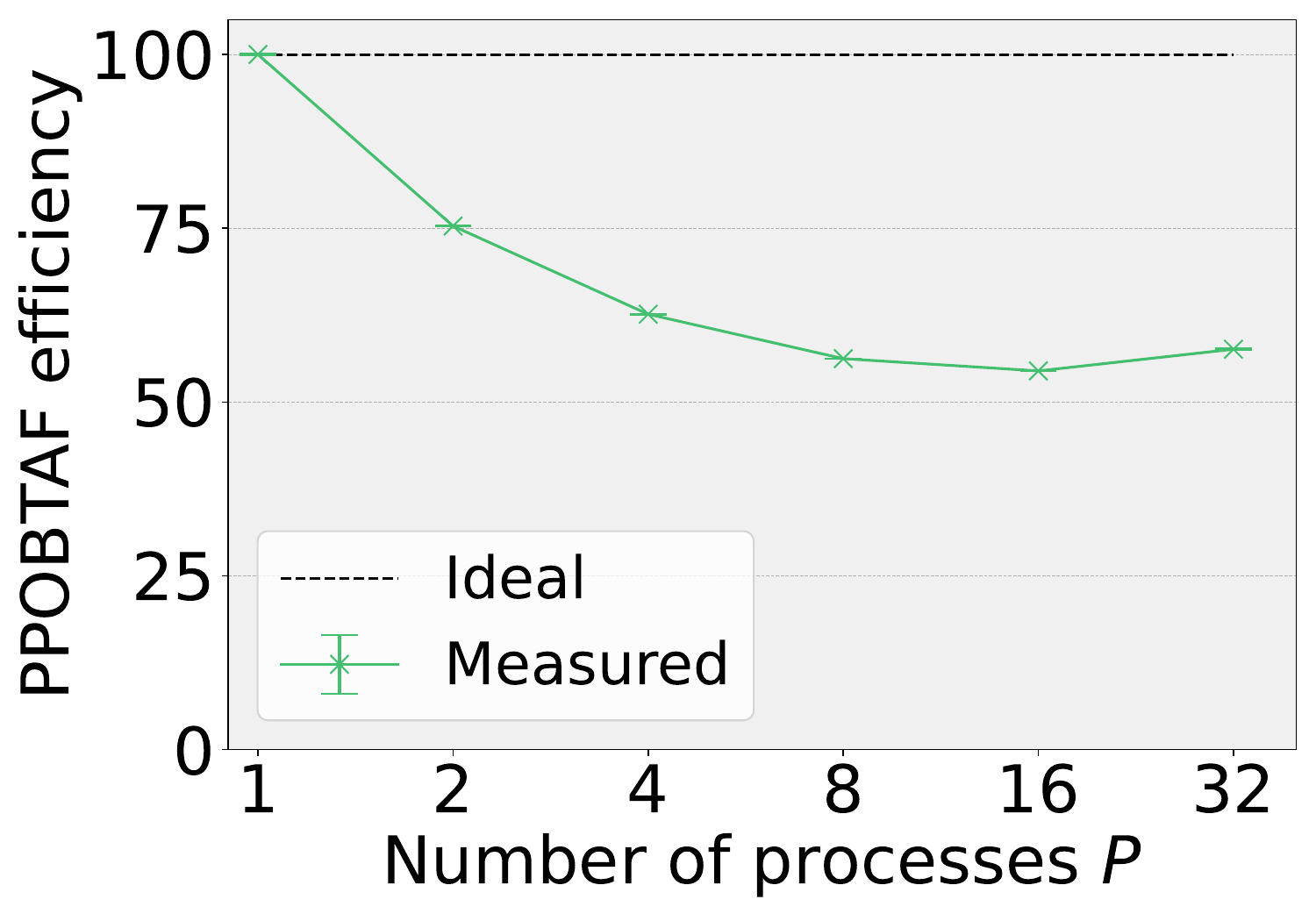}
    \end{subfigure}
    \begin{subfigure}[t]{.32\textwidth}
        \caption{}
        \includegraphics[width=\textwidth]{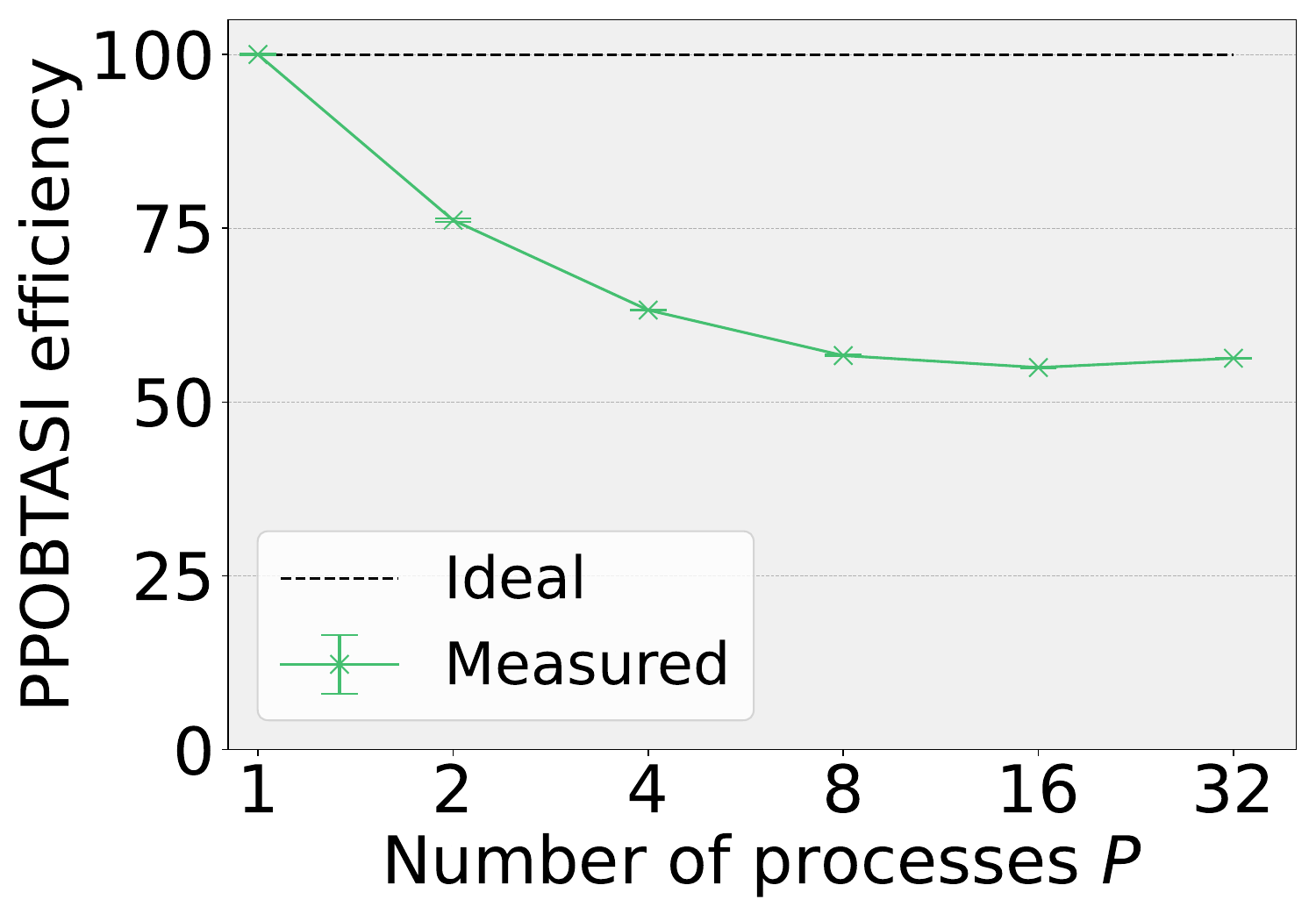}
    \end{subfigure}
    \begin{subfigure}[t]{.32\textwidth}
        \caption{}
        \includegraphics[width=\textwidth]{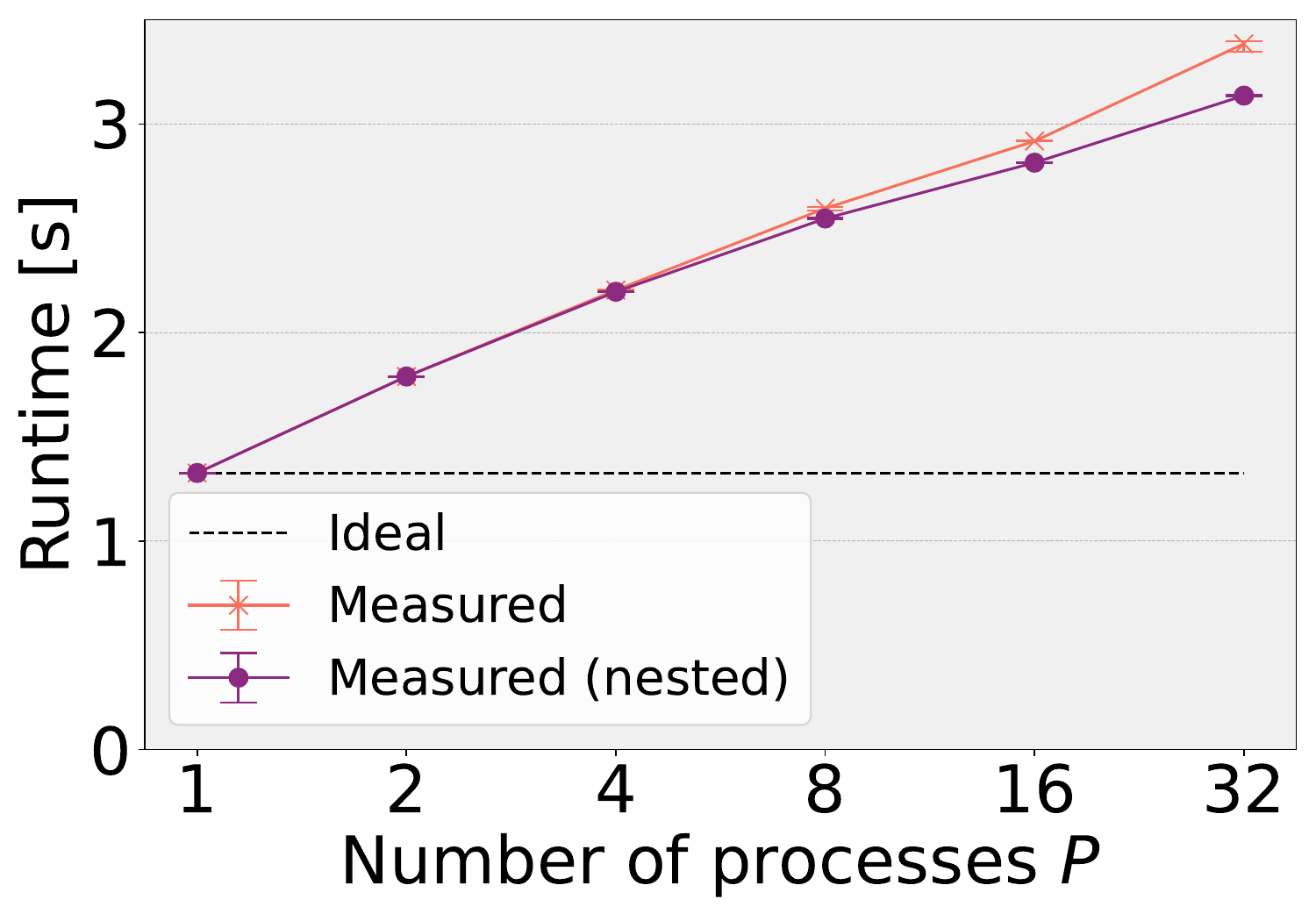}
    \end{subfigure}
    \caption{Weak-scaling results for the parallel selected-inversion algorithm on the Alps supercomputer from 1 to 32 processes (GPUs). a) Efficiency obtained for the PPOBTAF routine (parallel Cholesky factorization). b) Same as a), but for the PPOBTASI function (parallel selected inversion). c) Time-to-solution of the complete parallel approach. The solid lines refer to actual measurements; the dashed lines represent the ideal efficiency and runtime behavior.}
    \label{fig:experimental_weak_scaling}
\end{figure*}

\subsection{Experimental setup}
\label{sec:experimental-setup}
We perform experiments using four different BTA datasets.
First, we measure the parallel efficiency of our codes using a synthetic dataset, (1), that spans a wide range of problem sizes: $n \in \{32, 64, 128, 256, 512\}$, $b=1024$, and $a = 256$, giving an overview of the efficiency of our methods.
Then, we employ two datasets, (2) and (3), based on precision matrices arising from statistical modeling in air temperature prediction using the INLA method, specifically: $n=365$, $b=2865$, $a=4$ and $n=250$, $b=4002$, $a=6$.
Dataset (2) is presented in 3 different BTA densities, representing different mesh conectivities: $d \in \{0.69\%, 3\%, 5\%\}$.
Dataset (3) presents a BTA-density $d=0.52\%$.
The overall densities of these sparse matrices vary between $0.006\%$ and $0.041\%$.
See Fig.~\ref{fig:spy_bta} for a visual representation.
Finally, we construct a fourth dataset (4) to perform a weak scaling analysis.
This dataset is based on the same sparse precision matrix as dataset (2), but here we devise a per-process temporal discretization over one quarter of a year, i.e., 90 days, so that the partition size is $n_p=90$ before considering load balancing.

More precisely, each dataset contains the following BTA matrices:
\begin{enumerate}
    \item $n \in \{32, 64, 128, 256, 512\}, b=1024, a=256$ 
    \item $n=365, b=2865, a=4$, $d \in \{0.69\%,3\%,5\%\} $, 
    \item  $n=250, b=4002, a=6$, $d=0.52\%$
    \item $n \in \{90, 180, 360, 720, 1440, 2880\}, b=2865 ,a=4$
\end{enumerate}

We perform our CPU experiments on the Fritz cluster of the Erlangen National High Performance Computing Center (NHR@FAU).
We utilize the spr1tb partition, where each node is equipped with 2 52-core Intel Xeon Platinum 8470 (Sapphire Rapids) CPUs.
The nodes are connected to an Infiniband network.
We run our GPU experiments on the Alps supercomputer of the Swiss National Supercomputing Center (CSCS).
Each compute node comprises four NVIDIA GH200 Superchips.

The Grace CPU has 72 Arm Neoverse V2 cores, while the Hopper GPU, on which our code runs, has a maximum theoretical performance of 34 TFLOPS with the regular double-precision floating-point units and 67 TFLOPS when using the double-precision tensor cores.
Each Superchip has 128GB LPDDR5X and 96GB HMB3 memory.

All nodes are connected using an HPC Cray Slingshot-11 network with 200 Gbps injection bandwidth per Superchip.
Intra-Superchip and Inter-GPU (up to 256 GPUs) communication relies on NVLink with 900GB/s bandwidth.
All operations are done in double precision.
All runtimes presented are median values out of 10 executions, while the error bars correspond to the 95\% confidence intervals.

\subsection{Experimental parallel efficiency}
\label{sec:experimental-efficiency}

Using dataset (1), we measure the experimental parallel efficiency of the entire distributed selected-inversion procedure for different $P$ and $n$ values, but constant $b=1024$ and $a=256$, and present it in Fig.~\ref{fig:theoretical_perf_matrix}.b).
We compute the efficiency as the ratio of parallel runtime to the block-sequential execution.
The runtimes in parentheses are median values out of 20 measurements.

We make the following observations with respect to the theoretical efficiency in Fig.~\ref{fig:theoretical_perf_matrix}.a).
The two-process efficiency is about 10\% lower than expected.
This difference could be attributed to the fact that communication is not considered in our theoretical analysis.
However, we do not observe such discrepancies across the board.
% Another potential culprit is sub-optimal load balancing.
As we run our algorithms using the theoretical load balancing computed in \ref{tab:load_balancing}, by profiling the execution, we find that for $P=2$ and $n=512$, the PPOBTAF runtime is 475 ms, and the middle process spends about 180 ms waiting for the top process, implying that the theoretically determined ideal load balancing factor of $2.25$ is sub-optimal.
We further discuss this sub-optimal load balancing in the following paragraph.

\subsubsection{Discussion on the load balancing}
\label{sec:discussion}
The experimental results differ from the theoretical ones, partially due to sub-optimal load balancing.
Here, we reiterate that in typical scientific problems, $a$ and $b$ do not grow arbitrarily large.
They take instead relatively small values, which pose a challenge in predicting the performance of BLAS- and LAPACK-based implementations because matrix-matrix multiplication exposes higher parallelism than triangular routines such as POTRF and TRSM.
This issue is further exacerbated in state-of-the-art GPU accelerators, where vendor-optimized GEMM implementations may use specialized matrix-product units (tensor cores) efficiently.
To provide a concrete example, we show in Fig.~\ref{fig:sequential_cost_breakdown}.a) and b) the theoretical FLOP count distribution of the GEMM, TRSM, and POTRF routines for POBTAF and POBTASI, using $b=1024$ and $a=256$.
Both methods appear to be dominated by the GEMM operations.
However, when we measure the actual performance, we observe that most of the runtime is spent on TRSM and POTRF, which perform poorly relative to GEMM.
The kernels' runtime breakdown and performance in TFLOPS on GH200's GPU are given in Fig.~\ref{fig:sequential_cost_breakdown}.c) and d).
POTRF's runtime ratio in POBTAF is 32.5\%.
Since increasing the top partition's workload translates into increased POTRF calls, the theoretically optimal load balancing ratio may differ significantly from the practically optimal one.
For these reasons, we decided to fine-tune the load balancing factor that we use in Section~\ref{sec:soa_compare} with respect to the given problem and hardware.
To conclude, although fine-tuning the load-balancing ratio to specific problem sizes is, of course, feasible, it would be preferable to identify means to automatically derive better ratios for any parameters. It poses a direction for future work.

\begin{figure*}[t]
    \centering
    \includegraphics[width=.95\textwidth]{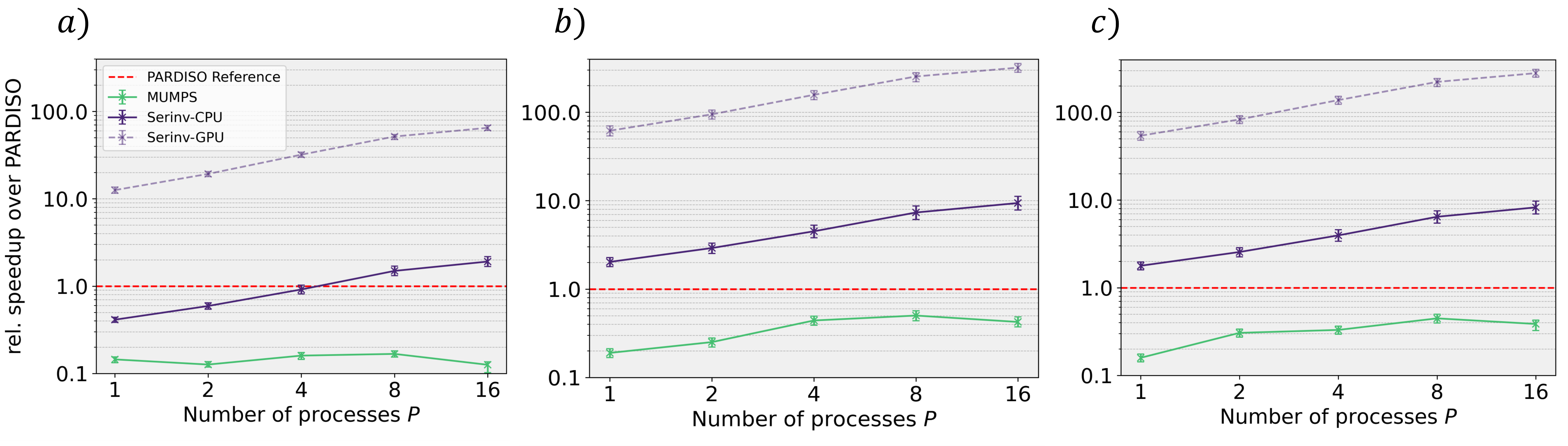}
    \caption{Relative speedup for selected matrix inversion (including factorization phase) of MUMPS, Serinv-CPU, and Serinv-GPU using an increasing number of processes over the best multithreaded PARDISO configuration (32 threads). 
    The dataset (2) and its BTA-density declinations are used, leading from left to right: a) Dataset (2), with $0.7\%$ BTA-density. b) Dataset (2), with $3\%$ BTA-density. c) Dataset (2), with $5\%$ BTA-density. CPU codes run on the Fritz cluster and Serinv-GPU on the Alps supercomputer.}
    \label{fig:relative_speedup_case1}
\end{figure*}

\subsection{Weak scaling}
\label{sec:scaling}

We perform weak-scaling experiments on the Alps supercomputer using dataset (4) and present the results in Fig.~\ref{fig:experimental_weak_scaling}.
The left (middle) subfigures show PPOBTAF's (PPOBTASI's) speedup and efficiency.
The parallel methods exhibit almost perfect scaling, except for an efficiency loss from 1 to 2 processes.
The block-sequential POBTAF runs for 2.51s, while PPOBTAF's execution with two processes takes 1.56s, of which about 120ms is the latency of an Allreduce operation with NCCL.
Furthermore, the middle process spends 200ms waiting for the top process, pointing again to sub-optimal load balancing.
We present in the right subfigure the total runtime for the parallel selected-inversion operation with and without nested solving.
As, with 90 diagonal blocks per process, dataset (4) is large, the nested solving's communication overhead is relatively small compared to the added computational cost, leading to a performance improvement for 16 and 32 processes.
The observed parallel efficiency is 47.2\% and 42.3\% at 16 and 32 GPUs, respectively.
As datasets (1-3) are much smaller, the nested solving approach does not bring any performance gains and, therefore, we do not present it further.

\begin{figure*}[t]
    \centering
    \includegraphics[width=.95\textwidth]{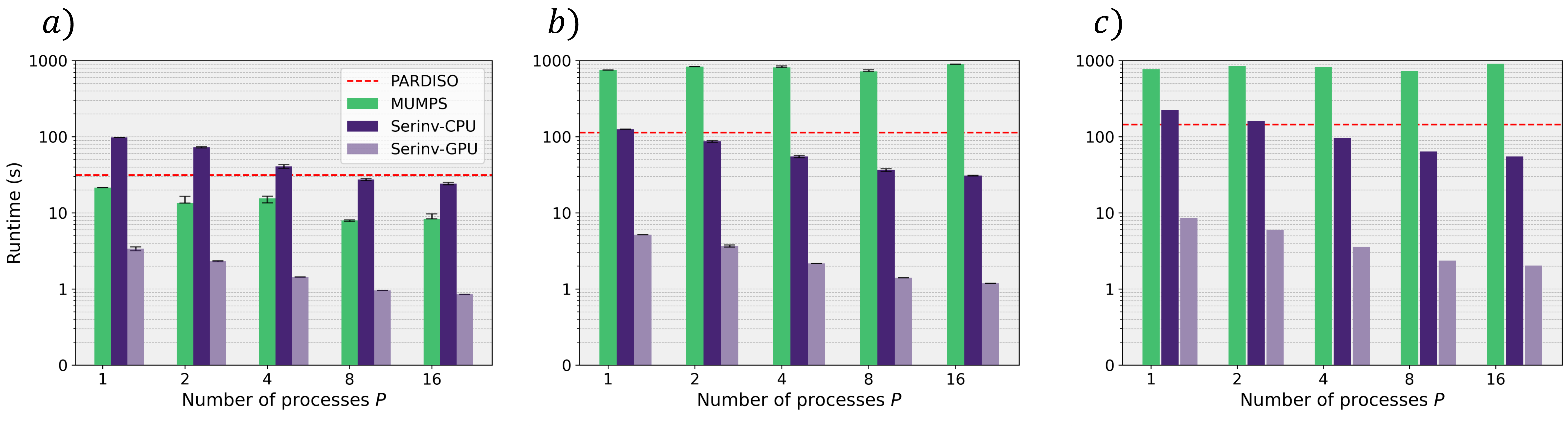}
    \caption{Runtime breakdown in Factorization and Selected Inversion of PARDISO and strong-scaling of MUMPS, Serinv-CPU, and Serinv-GPU with an increasing number of nodes over the best multithreaded PARDISO configuration (32 threads). a) factorization runtime, b) selected inversion runtime, and c) total time-to-solution a) + b). CPU codes run on the Fritz cluster and Serinv-GPU on the Alps supercomputer.}
    \label{fig:barchart_comparison}
\end{figure*}

\subsection{Comparison to State-of-the-Art}\label{sec:soa_compare}

We use datasets (2) and (3) to compare Serinv performances against the state-of-the-art sparse direct solvers PARDISO (version 7.2) and MUMPS (version 5.7.3). 
To provide a fair comparison, the sparse solvers utilize the true underlying sparsity pattern of the matrix and not its block-dense version.
% We compare the performance of Serinv to the software libraries PARDISO and MUMPS. 
Both libraries provide factorization and selected matrix inversion routines for general sparse matrices and, therefore, also support BTA sparsity patterns. 
These libraries rely on METIS/ParMETIS~\cite{karypis1998fast} for providing suitable matrix permutations to minimize fill-in during the factorization. 
Their complete selected inversion routines can, therefore, be split into three distinct phases: symbolic factorization, numerical factorization, and selected inversion. 
The first phase only needs to be performed once per sparsity pattern, allowing the amortization of its cost when for matrices with recurrent sparsity patterns. We, therefore, omit the inclusion of the analysis phase in the reported runtimes. 
The PARDISO library supports shared-memory parallelism, while MUMPS also offers a distributed-memory implementation.
However, neither of them offers GPU implementations of their sparse routines.
After testing the libraries' performances, we found the best configuration for each of them to be: 32 OpenMP/MKL threads for PARDISO and 52 for MUMPS and Serinv-CPU.
Furthermore, following MUMPS's documentation~\cite{mumps_doc}, we found that MUMPS performs best when operating on blocks of 2048 columns at a time, and we set the ICNTL(27) parameter appropriately.

As discussed in Section~\ref{sec:discussion}, we experimentally adjust the load balancing factor used in the multi-process runs of Serinv, where we found a ratio of about $1.8$ to be optimal.
We found PARDISO to be the fastest available library in literature, and thus present in Figure~\ref{fig:relative_speedup_case1} the strong-scaling speedup of MUMPS and Serinv relative to the best performance achieved by PARDISO.
The speedup results presented in Figure~\ref{fig:relative_speedup_case1} encompass the factorization and selected inversion phases of the matrix from dataset (2) and its different BTA-density declination.
For all three variations of BTA-density, Serinv-CPU (resp. Serinv-GPU) exhibits the same runtime, as the in-block density doesn't affect the operations performed.
The sparse implementation of PARDISO can leverage lower densities to its advantage, outperforming Serinv-CPU's single-node version.
We can, however, see from Figure~\ref{fig:relative_speedup_case1} that already from a BTA-density of $3\%$, Serinv's single node CPU implementation outperforms PARDISO, leveraging its adaption to the BTA sparsity pattern. 
Serinv-CPU (resp. GPU) achieves a strong scaling parallel efficiency of 28.9\% (resp. 32.9\%) on dataset (2) initial matrix, leading to a speedup of 4.63 (resp. 5.17) over its single-process implementation.

In Fig.~\ref{fig:barchart_comparison}, we present a breakdown of the runtimes for the factorization and selected inversion of the matrix from dataset (3), as well as their combined runtimes. 
We observe that MUMPS's factorization routine is performant and scalable, outperforming Serinv-CPU and PARDISO for any number of processes.
However, MUMPS' selected inversion takes about an order of magnitude longer than the two other CPU variants, making it the least-suited method to perform selected inversion on sparse BTA matrices. 
% As for both sparse solvers the selected inversion
% PARDISO's selected inversion takes about 3-4 longer than its factorization.
Both sparse solvers outperform Serinv-CPU during the factorization phase. 
Conversely, for the selected inversion routine, Serinv-CPU outperforms MUMPS even on a single process and PARDISO from 2 processes on.
The combined runtimes make PARDISO, up to 4 MPI processes, the fastest CPU solver.
Serinv-CPU scales further to 16 MPI processes, providing the best overall time-to-solution for a CPU solver with $55.09$ seconds, PARDISO (resp. MUMPS) taking $145.14$ (resp. $729.2$) seconds.
The single-process GPU implementation of Serinv outperforms all CPU variants for both the factorization and selected inversion phases, achieving $4.62$ times speedup over MUMPS' factorization and $21.95$ times speedup over PARDISO' selected inversion.
Serinv-GPU provides, in all cases, the fastest factorization, selected inversion, and time-to-solution.

We conclude that even for very sparse BTA matrices, Serinv-CPU already provides comparable performance to PARDISO. 
When increasing the BTA density above $3\%$, Serinv-CPU becomes the most efficient option.
Serinv-GPU outperforms all CPU solvers by one to two orders of magnitudes depending on the configuration ($71.4$ times speedup over PARDISO and $380.9$ times speedup over MUMPS on dataset (3)).

\section{Conclusion}\label{sec:conc} % 1 column

We derived novel parallel algorithms for the selected inversion of positive semi-definite BTA matrices and implemented them in the scalable, GPU-accelerated \textit{Serinv} library.
We conducted a theoretical analysis of the proposed methods and evaluated our implementations on synthetics and real datasets.
We first demonstrated weak scaling efficiency of $47.2\%$ when going from 1 to 16 GPUs.
We then conducted a comparison of Serinv's performances against the state-of-the-art sparse solvers PARDISO and MUMPS for which we show up to 2.6x (resp. 71.4x) speedup on CPU backend (resp. GPU) over PARDISO and 14.0x (resp. 380.9x) speedup over MUMPS when scaling to 16 processes (resp. GPUs).
Future work may include: (i) further theoretical study, especially regarding nested solving, to uncover better scaling opportunities; (ii) usage of heuristics to quickly determine better load-balancing ratios tailored to the exact problem parameters; and (iii) investigation of custom kernels that combine executions of POBTAF, GEMM, and TRSM to accelerate the intra-device execution.
Finally, the proposed distributed selected-inversion algorithms have the ability to extend the feasible scale of important materials science and climate modeling applications on state-of-the-art supercomputers, enabling new scientific discoveries.

\begin{acks}
This work received funding from the Swiss National Science Foundation (SNSF) under the grant agreements “Quantum Transport Simulations at the Exascale and Beyond (QuaTrEx)”
(n◦ 209358). We acknowledge support and HPC resources provided by the Swiss National Supercomputing Center (CSCS) under project lp16 as well as the Erlangen National High Performance Computing Center (NHR@FAU) of the Friedrich-Alexander-Universität Erlangen-Nürnberg (FAU) under the NHR project 80227.
\end{acks}

\bibliographystyle{ACM-Reference-Format}
\bibliography{bibl_conf}


\begin{thebibliography}{51}

%%% the Web address.

\ifx \showCODEN    \undefined \def \showCODEN     #1{\unskip}     \fi
\ifx \showDOI      \undefined \def \showDOI       #1{#1}\fi
\ifx \showISBNx    \undefined \def \showISBNx     #1{\unskip}     \fi
\ifx \showISBNxiii \undefined \def \showISBNxiii  #1{\unskip}     \fi
\ifx \showISSN     \undefined \def \showISSN      #1{\unskip}     \fi
\ifx \showLCCN     \undefined \def \showLCCN      #1{\unskip}     \fi
\ifx \shownote     \undefined \def \shownote      #1{#1}          \fi
\ifx \showarticletitle \undefined \def \showarticletitle #1{#1}   \fi
\ifx \showURL      \undefined \def \showURL       {\relax}        \fi
% invisible to TeX
\providecommand\bibfield[2]{#2}
\providecommand\bibinfo[2]{#2}
\providecommand\natexlab[1]{#1}
\providecommand\showeprint[2][]{arXiv:#2}

\bibitem[mum(2024)]%
        {mumps_doc}
 \bibinfo{year}{2024}\natexlab{}.
\newblock \bibinfo{title}{{MUltifrontal Massively Parallel Solver (MUMPS 5.7.3) Users’ guide}}.
\newblock
\newblock
\urldef\tempurl%
\url{https://mumps-solver.org/doc/userguide_5.7.3.pdf}
\showURL{%
\tempurl}


\bibitem[ncc(2024a)]%
        {nccl_1}
 \bibinfo{year}{2024}\natexlab{a}.
\newblock \bibinfo{title}{{NVIDIA} {Collective} {Communication} {Library} ({NCCL}) {Documentation} — {NCCL} 2.23.4 documentation}.
\newblock
\newblock
\urldef\tempurl%
\url{https://docs.nvidia.com/deeplearning/nccl/user-guide/docs/index.html}
\showURL{%
\tempurl}


\bibitem[ncc(2024b)]%
        {nccl_3}
 \bibinfo{year}{2024}\natexlab{b}.
\newblock \bibinfo{title}{{NVIDIA}/nccl}.
\newblock
\newblock
\urldef\tempurl%
\url{https://github.com/NVIDIA/nccl}
\showURL{%
\tempurl}
\newblock
\shownote{original-date: 2015-11-14T00:12:04Z}.


\bibitem[Amestoy et~al\mbox{.}(2000)]%
        {mumps_1}
\bibfield{author}{\bibinfo{person}{P.~R. Amestoy}, \bibinfo{person}{I.~S. Duff}, {and} \bibinfo{person}{J.~Y. L'Excellent}.} \bibinfo{year}{2000}\natexlab{}.
\newblock \showarticletitle{Multifrontal parallel distributed symmetric and unsymmetric solvers}.
\newblock \bibinfo{journal}{\emph{Computer Methods in Applied Mechanics and Engineering}} \bibinfo{volume}{184}, \bibinfo{number}{2} (\bibinfo{date}{April} \bibinfo{year}{2000}), \bibinfo{pages}{501--520}.
\newblock
\showISSN{0045-7825}
\urldef\tempurl%
\url{https://doi.org/10.1016/S0045-7825(99)00242-X}
\showDOI{\tempurl}


\bibitem[Amestoy et~al\mbox{.}(2001a)]%
        {mumps_2}
\bibfield{author}{\bibinfo{person}{Patrick~R. Amestoy}, \bibinfo{person}{Iain~S. Duff}, \bibinfo{person}{Jean-Yves L'Excellent}, {and} \bibinfo{person}{Jacko Koster}.} \bibinfo{year}{2001}\natexlab{a}.
\newblock \showarticletitle{A {Fully} {Asynchronous} {Multifrontal} {Solver} {Using} {Distributed} {Dynamic} {Scheduling}}.
\newblock \bibinfo{journal}{\emph{SIAM J. Matrix Anal. Appl.}} \bibinfo{volume}{23}, \bibinfo{number}{1} (\bibinfo{date}{Jan.} \bibinfo{year}{2001}), \bibinfo{pages}{15--41}.
\newblock
\showISSN{0895-4798}
\urldef\tempurl%
\url{https://doi.org/10.1137/S0895479899358194}
\showDOI{\tempurl}
\newblock
\shownote{Publisher: Society for Industrial and Applied Mathematics}.


\bibitem[Amestoy et~al\mbox{.}(2001b)]%
        {mumps_3}
\bibfield{author}{\bibinfo{person}{Patrick~R. Amestoy}, \bibinfo{person}{Iain~S. Duff}, \bibinfo{person}{Jean-Yves L’Excellent}, {and} \bibinfo{person}{Jacko Koster}.} \bibinfo{year}{2001}\natexlab{b}.
\newblock \showarticletitle{{MUMPS}: {A} {General} {Purpose} {Distributed} {Memory} {Sparse} {Solver}}. In \bibinfo{booktitle}{\emph{Applied {Parallel} {Computing}. {New} {Paradigms} for {HPC} in {Industry} and {Academia}}}, \bibfield{editor}{\bibinfo{person}{Tor Sørevik}, \bibinfo{person}{Fredrik Manne}, \bibinfo{person}{Assefaw~Hadish Gebremedhin}, {and} \bibinfo{person}{Randi Moe}} (Eds.). \bibinfo{publisher}{Springer}, \bibinfo{address}{Berlin, Heidelberg}, \bibinfo{pages}{121--130}.
\newblock
\showISBNx{978-3-540-70734-9}
\urldef\tempurl%
\url{https://doi.org/10.1007/3-540-70734-4_16}
\showDOI{\tempurl}


\bibitem[Awan et~al\mbox{.}(2018)]%
        {nccl_4}
\bibfield{author}{\bibinfo{person}{Ammar~Ahmad Awan}, \bibinfo{person}{Ching-Hsiang Chu}, \bibinfo{person}{Hari Subramoni}, {and} \bibinfo{person}{Dhabaleswar~K. Panda}.} \bibinfo{year}{2018}\natexlab{}.
\newblock \showarticletitle{Optimized Broadcast for Deep Learning Workloads on Dense-GPU InfiniBand Clusters: MPI or NCCL?}. In \bibinfo{booktitle}{\emph{Proceedings of the 25th European MPI Users' Group Meeting}} (Barcelona, Spain) \emph{(\bibinfo{series}{EuroMPI '18})}. \bibinfo{publisher}{Association for Computing Machinery}, \bibinfo{address}{New York, NY, USA}, Article \bibinfo{articleno}{2}, \bibinfo{numpages}{9}~pages.
\newblock
\showISBNx{9781450364928}
\urldef\tempurl%
\url{https://doi.org/10.1145/3236367.3236381}
\showDOI{\tempurl}


\bibitem[Belov et~al\mbox{.}(2017)]%
        {arrowhead_decomp}
\bibfield{author}{\bibinfo{person}{P~A Belov}, \bibinfo{person}{E~R Nugumanov}, {and} \bibinfo{person}{S~L Yakovlev}.} \bibinfo{year}{2017}\natexlab{}.
\newblock \showarticletitle{The arrowhead decomposition method for a block-tridiagonal system of linear equations}.
\newblock \bibinfo{journal}{\emph{Journal of Physics: Conference Series}} \bibinfo{volume}{929}, \bibinfo{number}{1} (\bibinfo{date}{nov} \bibinfo{year}{2017}), \bibinfo{pages}{012035}.
\newblock
\urldef\tempurl%
\url{https://doi.org/10.1088/1742-6596/929/1/012035}
\showDOI{\tempurl}


\bibitem[Bollhöfer et~al\mbox{.}(2020)]%
        {pardiso_3}
\bibfield{author}{\bibinfo{person}{Matthias Bollhöfer}, \bibinfo{person}{Olaf Schenk}, \bibinfo{person}{Radim Janalik}, \bibinfo{person}{Steve Hamm}, {and} \bibinfo{person}{Kiran Gullapalli}.} \bibinfo{year}{2020}\natexlab{}.
\newblock \showarticletitle{State-of-the-{Art} {Sparse} {Direct} {Solvers}}.
\newblock In \bibinfo{booktitle}{\emph{Parallel {Algorithms} in {Computational} {Science} and {Engineering}}}, \bibfield{editor}{\bibinfo{person}{Ananth Grama} {and} \bibinfo{person}{Ahmed~H. Sameh}} (Eds.). \bibinfo{publisher}{Springer International Publishing}, \bibinfo{address}{Cham}, \bibinfo{pages}{3--33}.
\newblock
\showISBNx{978-3-030-43736-7}
\urldef\tempurl%
\url{https://doi.org/10.1007/978-3-030-43736-7\_1}
\showDOI{\tempurl}


\bibitem[Bondeli and Gander(1994)]%
        {bcr_2}
\bibfield{author}{\bibinfo{person}{S. Bondeli} {and} \bibinfo{person}{W. Gander}.} \bibinfo{year}{1994}\natexlab{}.
\newblock \showarticletitle{Cyclic {Reduction} for {Special} {Tridiagonal} {Systems}}.
\newblock \bibinfo{journal}{\emph{SIAM J. Matrix Anal. Appl.}} \bibinfo{volume}{15}, \bibinfo{number}{1} (\bibinfo{date}{Jan.} \bibinfo{year}{1994}), \bibinfo{pages}{321--330}.
\newblock
\showISSN{0895-4798}
\urldef\tempurl%
\url{https://doi.org/10.1137/S0895479891220533}
\showDOI{\tempurl}
\newblock
\shownote{Publisher: Society for Industrial and Applied Mathematics}.


\bibitem[Calderara(2016)]%
        {splitsolve}
\bibfield{author}{\bibinfo{person}{Mauro~M. Calderara}.} \bibinfo{year}{2016}\natexlab{}.
\newblock \emph{\bibinfo{title}{{SplitSolve}, an {Algorithm} for {Ab}-{Initio} {Quantum} {Transport} {Simulations}}}.
\newblock Doctoral {Thesis}. \bibinfo{school}{ETH Zurich}.
\newblock
\urldef\tempurl%
\url{https://doi.org/10.3929/ethz-a-010781848}
\showDOI{\tempurl}
\newblock
\shownote{Accepted: 2017-09-22T13:18:14Z}.


\bibitem[Cauley et~al\mbox{.}(2007)]%
        {pdiv_2}
\bibfield{author}{\bibinfo{person}{Stephen Cauley}, \bibinfo{person}{Jitesh Jain}, \bibinfo{person}{Cheng-Kok Koh}, {and} \bibinfo{person}{Venkataramanan Balakrishnan}.} \bibinfo{year}{2007}\natexlab{}.
\newblock \showarticletitle{A scalable distributed method for quantum-scale device simulation}.
\newblock \bibinfo{journal}{\emph{Journal of Applied Physics}} \bibinfo{volume}{101}, \bibinfo{number}{12} (\bibinfo{date}{June} \bibinfo{year}{2007}), \bibinfo{pages}{123715}.
\newblock
\showISSN{0021-8979}
\urldef\tempurl%
\url{https://doi.org/10.1063/1.2748621}
\showDOI{\tempurl}


\bibitem[Cauley et~al\mbox{.}(2011)]%
        {pdiv_3}
\bibfield{author}{\bibinfo{person}{Stephen Cauley}, \bibinfo{person}{Mathieu Luisier}, \bibinfo{person}{Venkataramanan Balakrishnan}, \bibinfo{person}{Gerhard Klimeck}, {and} \bibinfo{person}{Cheng-Kok Koh}.} \bibinfo{year}{2011}\natexlab{}.
\newblock \showarticletitle{Distributed non-equilibrium {Green}’s function algorithms for the simulation of nanoelectronic devices with scattering}.
\newblock \bibinfo{journal}{\emph{Journal of Applied Physics}} \bibinfo{volume}{110}, \bibinfo{number}{4} (\bibinfo{date}{Aug.} \bibinfo{year}{2011}), \bibinfo{pages}{043713}.
\newblock
\showISSN{0021-8979}
\urldef\tempurl%
\url{https://doi.org/10.1063/1.3624612}
\showDOI{\tempurl}


\bibitem[Dalcín et~al\mbox{.}(2005)]%
        {mpi4py}
\bibfield{author}{\bibinfo{person}{Lisandro Dalcín}, \bibinfo{person}{Rodrigo Paz}, {and} \bibinfo{person}{Mario Storti}.} \bibinfo{year}{2005}\natexlab{}.
\newblock \showarticletitle{MPI for Python}.
\newblock \bibinfo{journal}{\emph{J. Parallel and Distrib. Comput.}} \bibinfo{volume}{65}, \bibinfo{number}{9} (\bibinfo{year}{2005}), \bibinfo{pages}{1108--1115}.
\newblock
\showISSN{0743-7315}
\urldef\tempurl%
\url{https://doi.org/10.1016/j.jpdc.2005.03.010}
\showDOI{\tempurl}


\bibitem[Demmel et~al\mbox{.}(1995)]%
        {block-lu-demmel}
\bibfield{author}{\bibinfo{person}{James~W. Demmel}, \bibinfo{person}{Nicholas~J. Higham}, {and} \bibinfo{person}{Robert~S. Schreiber}.} \bibinfo{year}{1995}\natexlab{}.
\newblock \showarticletitle{Stability of block LU factorization}.
\newblock \bibinfo{journal}{\emph{Numerical Linear Algebra with Applications}} \bibinfo{volume}{2}, \bibinfo{number}{2} (\bibinfo{year}{1995}), \bibinfo{pages}{173--190}.
\newblock
\urldef\tempurl%
\url{https://doi.org/10.1002/nla.1680020208}
\showDOI{\tempurl}
\showeprint{https://onlinelibrary.wiley.com/doi/pdf/10.1002/nla.1680020208}


\bibitem[Erisman and Tinney(1975)]%
        {erisman1975computing}
\bibfield{author}{\bibinfo{person}{AM Erisman} {and} \bibinfo{person}{WF Tinney}.} \bibinfo{year}{1975}\natexlab{}.
\newblock \showarticletitle{On computing certain elements of the inverse of a sparse matrix}.
\newblock \bibinfo{journal}{\emph{Commun. ACM}} \bibinfo{volume}{18}, \bibinfo{number}{3} (\bibinfo{year}{1975}), \bibinfo{pages}{177--179}.
\newblock


\bibitem[Gaedke-Merzh\"{a}user(2024)]%
        {inla_dist}
\bibfield{author}{\bibinfo{person}{Lisa Gaedke-Merzh\"{a}user}.} \bibinfo{year}{2024}\natexlab{}.
\newblock \bibinfo{title}{lisa-gm/{INLA}\_DIST}.
\newblock
\newblock
\urldef\tempurl%
\url{https://github.com/lisa-gm/INLA\_DIST}
\showURL{%
\tempurl}
\newblock
\shownote{original-date: 2021-04-14T07:35:36Z}.


\bibitem[Gaedke-Merzh\"{a}user et~al\mbox{.}(2024)]%
        {gaedkeIntegrated2024}
\bibfield{author}{\bibinfo{person}{Lisa Gaedke-Merzh\"{a}user}, \bibinfo{person}{Elias Krainski}, \bibinfo{person}{Radim Janalik}, \bibinfo{person}{H\r{a}vard Rue}, {and} \bibinfo{person}{Olaf Schenk}.} \bibinfo{year}{2024}\natexlab{}.
\newblock \showarticletitle{{Integrated Nested Laplace Approximations for Large-Scale Spatiotemporal Bayesian Modeling}}.
\newblock \bibinfo{journal}{\emph{SIAM Journal on Scientific Computing}} \bibinfo{volume}{46}, \bibinfo{number}{4} (\bibinfo{year}{2024}), \bibinfo{pages}{B448--B473}.
\newblock
\urldef\tempurl%
\url{https://doi.org/10.1137/23M1561531}
\showDOI{\tempurl}


\bibitem[Gaedke-Merzh{\"a}user et~al\mbox{.}(2023)]%
        {gaedke2022parallelized}
\bibfield{author}{\bibinfo{person}{Lisa Gaedke-Merzh{\"a}user}, \bibinfo{person}{Janet van Niekerk}, \bibinfo{person}{Olaf Schenk}, {and} \bibinfo{person}{H{\aa}vard Rue}.} \bibinfo{year}{2023}\natexlab{}.
\newblock \showarticletitle{{Parallelized integrated nested {L}}aplace approximations for fast {Bayesian} inference}.
\newblock \bibinfo{journal}{\emph{Statistics and Computing}} \bibinfo{volume}{33}, \bibinfo{number}{25} (\bibinfo{year}{2023}).
\newblock
\newblock
\shownote{\href{https://doi.org/10.1007/s11222-022-10192-1}{{10.1007/s11222-022-10192-1}}}.


\bibitem[George et~al\mbox{.}(1987)]%
        {symbolic_cholesky}
\bibfield{author}{\bibinfo{person}{Alan George}, \bibinfo{person}{Micheal~T Heath}, \bibinfo{person}{Esmond Ng}, {and} \bibinfo{person}{Joseph Liu}.} \bibinfo{year}{1987}\natexlab{}.
\newblock \showarticletitle{Symbolic Cholesky factorization on a local-memory multiprocessor}.
\newblock \bibinfo{journal}{\emph{Parallel Comput.}} \bibinfo{volume}{5}, \bibinfo{number}{1} (\bibinfo{year}{1987}), \bibinfo{pages}{85--95}.
\newblock
\showISSN{0167-8191}
\urldef\tempurl%
\url{https://doi.org/10.1016/0167-8191(87)90009-3}
\showDOI{\tempurl}
\newblock
\shownote{Proceedings of the International Conference on Vector and Parallel Computing-Issues in Applied Research and Development}.


\bibitem[Gianinazzi et~al\mbox{.}(2024)]%
        {lukas}
\bibfield{author}{\bibinfo{person}{Lukas Gianinazzi}, \bibinfo{person}{Alexandros~Nikolaos Ziogas}, \bibinfo{person}{Langwen Huang}, \bibinfo{person}{Piotr Luczynski}, \bibinfo{person}{Saleh Ashkboosh}, \bibinfo{person}{Florian Scheidl}, \bibinfo{person}{Armon Carigiet}, \bibinfo{person}{Chio Ge}, \bibinfo{person}{Nabil Abubaker}, \bibinfo{person}{Maciej Besta}, \bibinfo{person}{Tal Ben-Nun}, {and} \bibinfo{person}{Torsten Hoefler}.} \bibinfo{year}{2024}\natexlab{}.
\newblock \showarticletitle{Arrow Matrix Decomposition: A Novel Approach for Communication-Efficient Sparse Matrix Multiplication}. In \bibinfo{booktitle}{\emph{Proceedings of the 29th ACM SIGPLAN Annual Symposium on Principles and Practice of Parallel Programming}} (Edinburgh, United Kingdom) \emph{(\bibinfo{series}{PPoPP '24})}. \bibinfo{publisher}{Association for Computing Machinery}, \bibinfo{address}{New York, NY, USA}, \bibinfo{pages}{404–416}.
\newblock
\showISBNx{9798400704352}
\urldef\tempurl%
\url{https://doi.org/10.1145/3627535.3638496}
\showDOI{\tempurl}


\bibitem[Harris et~al\mbox{.}(2020)]%
        {harris2020array}
\bibfield{author}{\bibinfo{person}{Charles~R. Harris}, \bibinfo{person}{K.~Jarrod Millman}, \bibinfo{person}{St{\'{e}}fan~J. van~der Walt}, \bibinfo{person}{Ralf Gommers}, \bibinfo{person}{Pauli Virtanen}, \bibinfo{person}{David Cournapeau}, \bibinfo{person}{Eric Wieser}, \bibinfo{person}{Julian Taylor}, \bibinfo{person}{Sebastian Berg}, \bibinfo{person}{Nathaniel~J. Smith}, \bibinfo{person}{Robert Kern}, \bibinfo{person}{Matti Picus}, \bibinfo{person}{Stephan Hoyer}, \bibinfo{person}{Marten~H. van Kerkwijk}, \bibinfo{person}{Matthew Brett}, \bibinfo{person}{Allan Haldane}, \bibinfo{person}{Jaime~Fern{\'{a}}ndez del R{\'{i}}o}, \bibinfo{person}{Mark Wiebe}, \bibinfo{person}{Pearu Peterson}, \bibinfo{person}{Pierre G{\'{e}}rard-Marchant}, \bibinfo{person}{Kevin Sheppard}, \bibinfo{person}{Tyler Reddy}, \bibinfo{person}{Warren Weckesser}, \bibinfo{person}{Hameer Abbasi}, \bibinfo{person}{Christoph Gohlke}, {and} \bibinfo{person}{Travis~E. Oliphant}.} \bibinfo{year}{2020}\natexlab{}.
\newblock \showarticletitle{Array programming with {NumPy}}.
\newblock \bibinfo{journal}{\emph{Nature}} \bibinfo{volume}{585}, \bibinfo{number}{7825} (\bibinfo{date}{Sept.} \bibinfo{year}{2020}), \bibinfo{pages}{357--362}.
\newblock
\urldef\tempurl%
\url{https://doi.org/10.1038/s41586-020-2649-2}
\showDOI{\tempurl}


\bibitem[Heller(1976)]%
        {bcr_1}
\bibfield{author}{\bibinfo{person}{Don Heller}.} \bibinfo{year}{1976}\natexlab{}.
\newblock \showarticletitle{Some {Aspects} of the {Cyclic} {Reduction} {Algorithm} for {Block} {Tridiagonal} {Linear} {Systems}}.
\newblock \bibinfo{journal}{\emph{SIAM J. Numer. Anal.}} \bibinfo{volume}{13}, \bibinfo{number}{4} (\bibinfo{year}{1976}), \bibinfo{pages}{484--496}.
\newblock
\showISSN{0036-1429}
\urldef\tempurl%
\url{https://www.jstor.org/stable/2156240}
\showURL{%
\tempurl}
\newblock
\shownote{Publisher: Society for Industrial and Applied Mathematics}.


\bibitem[K~Takahashi(1973)]%
        {takahashi_sellinv}
\bibfield{author}{\bibinfo{person}{Fagan~J K~Takahashi, Chin~M}.} \bibinfo{year}{1973}\natexlab{}.
\newblock \showarticletitle{Formation of sparse bus impedance matrix and its application to short circuit study}.
\newblock \bibinfo{journal}{\emph{Proceedings of the 8th PICA Conference}} (\bibinfo{year}{1973}).
\newblock


\bibitem[Karypis and Kumar(1998)]%
        {karypis1998fast}
\bibfield{author}{\bibinfo{person}{George Karypis} {and} \bibinfo{person}{Vipin Kumar}.} \bibinfo{year}{1998}\natexlab{}.
\newblock \showarticletitle{{A fast and high quality multilevel scheme for partitioning irregular graphs}}.
\newblock \bibinfo{journal}{\emph{SIAM Journal on scientific Computing}} \bibinfo{volume}{20}, \bibinfo{number}{1} (\bibinfo{year}{1998}), \bibinfo{pages}{359--392}.
\newblock
\newblock
\shownote{\href{https://dl.acm.org/doi/10.5555/305219.305248}{{10.5555/305219.305248}}}.


\bibitem[Li et~al\mbox{.}(2008)]%
        {find_1}
\bibfield{author}{\bibinfo{person}{S. Li}, \bibinfo{person}{S. Ahmed}, \bibinfo{person}{G. Klimeck}, {and} \bibinfo{person}{E. Darve}.} \bibinfo{year}{2008}\natexlab{}.
\newblock \showarticletitle{Computing entries of the inverse of a sparse matrix using the {FIND} algorithm}.
\newblock \bibinfo{journal}{\emph{J. Comput. Phys.}} \bibinfo{volume}{227}, \bibinfo{number}{22} (\bibinfo{date}{Nov.} \bibinfo{year}{2008}), \bibinfo{pages}{9408--9427}.
\newblock
\showISSN{0021-9991}
\urldef\tempurl%
\url{https://doi.org/10.1016/j.jcp.2008.06.033}
\showDOI{\tempurl}


\bibitem[Li and Darve(2009)]%
        {find_4}
\bibfield{author}{\bibinfo{person}{S. Li} {and} \bibinfo{person}{E. Darve}.} \bibinfo{year}{2009}\natexlab{}.
\newblock \showarticletitle{Optimization of the {FIND} {Algorithm} to {Compute} the {Inverse} of a {Sparse} {Matrix}}. In \bibinfo{booktitle}{\emph{2009 13th {International} {Workshop} on {Computational} {Electronics}}}. \bibinfo{pages}{1--4}.
\newblock
\urldef\tempurl%
\url{https://doi.org/10.1109/IWCE.2009.5091136}
\showDOI{\tempurl}


\bibitem[Li and Darve(2012)]%
        {find_3}
\bibfield{author}{\bibinfo{person}{S. Li} {and} \bibinfo{person}{E. Darve}.} \bibinfo{year}{2012}\natexlab{}.
\newblock \showarticletitle{Extension and optimization of the {FIND} algorithm: {Computing} {Green}’s and less-than {Green}’s functions}.
\newblock \bibinfo{journal}{\emph{J. Comput. Phys.}} \bibinfo{volume}{231}, \bibinfo{number}{4} (\bibinfo{date}{Feb.} \bibinfo{year}{2012}), \bibinfo{pages}{1121--1139}.
\newblock
\showISSN{0021-9991}
\urldef\tempurl%
\url{https://doi.org/10.1016/j.jcp.2011.05.027}
\showDOI{\tempurl}


\bibitem[Li et~al\mbox{.}(2013)]%
        {find_2}
\bibfield{author}{\bibinfo{person}{S. Li}, \bibinfo{person}{W. Wu}, {and} \bibinfo{person}{E. Darve}.} \bibinfo{year}{2013}\natexlab{}.
\newblock \showarticletitle{A fast algorithm for sparse matrix computations related to inversion}.
\newblock \bibinfo{journal}{\emph{J. Comput. Phys.}}  \bibinfo{volume}{242} (\bibinfo{date}{June} \bibinfo{year}{2013}), \bibinfo{pages}{915--945}.
\newblock
\showISSN{0021-9991}
\urldef\tempurl%
\url{https://doi.org/10.1016/j.jcp.2013.01.036}
\showDOI{\tempurl}


\bibitem[Lin et~al\mbox{.}(2009)]%
        {lin2009fast}
\bibfield{author}{\bibinfo{person}{Lin Lin}, \bibinfo{person}{Jianfeng Lu}, \bibinfo{person}{Lexing Ying}, \bibinfo{person}{Roberto Car}, {et~al\mbox{.}}} \bibinfo{year}{2009}\natexlab{}.
\newblock \showarticletitle{Fast algorithm for extracting the diagonal of the inverse matrix with application to the electronic structure analysis of metallic systems}.
\newblock \bibinfo{journal}{\emph{Commun. Math. Sci.}} \bibinfo{volume}{7}, \bibinfo{number}{1} (\bibinfo{year}{2009}), \bibinfo{pages}{755--777}.
\newblock


\bibitem[Lin et~al\mbox{.}(2011)]%
        {lin_selinv}
\bibfield{author}{\bibinfo{person}{Lin Lin}, \bibinfo{person}{Chao Yang}, \bibinfo{person}{Juan~C. Meza}, \bibinfo{person}{Jianfeng Lu}, \bibinfo{person}{Lexing Ying}, {and} \bibinfo{person}{Weinan E}.} \bibinfo{year}{2011}\natexlab{}.
\newblock \showarticletitle{SelInv---An Algorithm for Selected Inversion of a Sparse Symmetric Matrix}.
\newblock \bibinfo{journal}{\emph{ACM Trans. Math. Softw.}} \bibinfo{volume}{37}, \bibinfo{number}{4}, Article \bibinfo{articleno}{40} (\bibinfo{date}{feb} \bibinfo{year}{2011}), \bibinfo{numpages}{19}~pages.
\newblock
\showISSN{0098-3500}
\urldef\tempurl%
\url{https://doi.org/10.1145/1916461.1916464}
\showDOI{\tempurl}


\bibitem[Lindgren et~al\mbox{.}(2024)]%
        {lindgren2024diffusion}
\bibfield{author}{\bibinfo{person}{Finn Lindgren}, \bibinfo{person}{Haakon Bakka}, \bibinfo{person}{David Bolin}, \bibinfo{person}{Elias Krainski}, {and} \bibinfo{person}{H{\aa}vard Rue}.} \bibinfo{year}{2024}\natexlab{}.
\newblock \showarticletitle{A diffusion-based spatio-temporal extension of Gaussian Mat{\'e}rn fields}.
\newblock \bibinfo{journal}{\emph{SORT-Statistics and Operations Research Transactions}} (\bibinfo{year}{2024}), \bibinfo{pages}{3--66}.
\newblock


\bibitem[Lindgren et~al\mbox{.}(2011)]%
        {lindgren2011explicit}
\bibfield{author}{\bibinfo{person}{Finn Lindgren}, \bibinfo{person}{H{\aa}vard Rue}, {and} \bibinfo{person}{Johan Lindstr{\"o}m}.} \bibinfo{year}{2011}\natexlab{}.
\newblock \showarticletitle{{An explicit link between {G}}aussian fields and {G}aussian {M}arkov random fields: the stochastic partial differential equation approach}.
\newblock \bibinfo{journal}{\emph{{Journal of the Royal Statistical Society: Series B (Statistical Methodology)}}} \bibinfo{volume}{73}, \bibinfo{number}{4} (\bibinfo{year}{2011}), \bibinfo{pages}{423--498}.
\newblock
\newblock
\shownote{\href{https://doi.org/10.1111/j.1467-9868.2011.00777.x}{{10.1111/j.1467-9868.2011.00777.x}}}.


\bibitem[Louter-Nool(1992)]%
        {block-cholesky-louter-nool}
\bibfield{author}{\bibinfo{person}{M. Louter-Nool}.} \bibinfo{year}{1992}\natexlab{}.
\newblock \showarticletitle{Block-Cholesky for parallel processing}.
\newblock \bibinfo{journal}{\emph{Applied Numerical Mathematics}} \bibinfo{volume}{10}, \bibinfo{number}{1} (\bibinfo{date}{Jan.} \bibinfo{year}{1992}), \bibinfo{pages}{37--57}.
\newblock


\bibitem[Luisier and Klimeck(2008)]%
        {omen}
\bibfield{author}{\bibinfo{person}{Mathieu Luisier} {and} \bibinfo{person}{Gerhard Klimeck}.} \bibinfo{year}{2008}\natexlab{}.
\newblock \showarticletitle{OMEN an Atomistic and Full-Band Quantum Transport Simulator for post-CMOS Nanodevices}. In \bibinfo{booktitle}{\emph{2008 8th IEEE Conference on Nanotechnology}}. \bibinfo{pages}{354--357}.
\newblock
\urldef\tempurl%
\url{https://doi.org/10.1109/NANO.2008.110}
\showDOI{\tempurl}


\bibitem[Luisier et~al\mbox{.}(2006)]%
        {negf1}
\bibfield{author}{\bibinfo{person}{Mathieu Luisier}, \bibinfo{person}{Andreas Schenk}, {and} \bibinfo{person}{Wolfgang Fichtner}.} \bibinfo{year}{2006}\natexlab{}.
\newblock \showarticletitle{{Quantum transport in two- and three-dimensional nanoscale transistors: Coupled mode effects in the nonequilibrium Green’s function formalism}}.
\newblock \bibinfo{journal}{\emph{Journal of Applied Physics}} \bibinfo{volume}{100}, \bibinfo{number}{4} (\bibinfo{date}{08} \bibinfo{year}{2006}), \bibinfo{pages}{043713}.
\newblock
\showISSN{0021-8979}
\urldef\tempurl%
\url{https://doi.org/10.1063/1.2244522}
\showDOI{\tempurl}
\showeprint{https://pubs.aip.org/aip/jap/article-pdf/doi/10.1063/1.2244522/7892769/043713\_1\_online.pdf}


\bibitem[Okuta et~al\mbox{.}(2017)]%
        {cupy}
\bibfield{author}{\bibinfo{person}{Ryosuke Okuta}, \bibinfo{person}{Yuya Unno}, \bibinfo{person}{Daisuke Nishino}, \bibinfo{person}{Shohei Hido}, {and} \bibinfo{person}{Crissman Loomis}.} \bibinfo{year}{2017}\natexlab{}.
\newblock \showarticletitle{CuPy: A NumPy-Compatible Library for NVIDIA GPU Calculations}. In \bibinfo{booktitle}{\emph{Proceedings of Workshop on Machine Learning Systems (LearningSys) in The Thirty-first Annual Conference on Neural Information Processing Systems (NIPS)}}.
\newblock
\urldef\tempurl%
\url{http://learningsys.org/nips17/assets/papers/paper\_16.pdf}
\showURL{%
\tempurl}


\bibitem[Petersen et~al\mbox{.}(2009)]%
        {petersen_hybrid_2009}
\bibfield{author}{\bibinfo{person}{Dan~Erik Petersen}, \bibinfo{person}{Song Li}, \bibinfo{person}{Kurt Stokbro}, \bibinfo{person}{Hans Henrik~B. Sørensen}, \bibinfo{person}{Per~Christian Hansen}, \bibinfo{person}{Stig Skelboe}, {and} \bibinfo{person}{Eric Darve}.} \bibinfo{year}{2009}\natexlab{}.
\newblock \showarticletitle{A hybrid method for the parallel computation of {Green}’s functions}.
\newblock \bibinfo{journal}{\emph{J. Comput. Phys.}} \bibinfo{volume}{228}, \bibinfo{number}{14} (\bibinfo{date}{Aug.} \bibinfo{year}{2009}), \bibinfo{pages}{5020--5039}.
\newblock
\showISSN{0021-9991}
\urldef\tempurl%
\url{https://doi.org/10.1016/j.jcp.2009.03.035}
\showDOI{\tempurl}


\bibitem[Petersen et~al\mbox{.}(2008)]%
        {rgf_2}
\bibfield{author}{\bibinfo{person}{Dan~Erik Petersen}, \bibinfo{person}{Hans Henrik~B. Sørensen}, \bibinfo{person}{Per~Christian Hansen}, \bibinfo{person}{Stig Skelboe}, {and} \bibinfo{person}{Kurt Stokbro}.} \bibinfo{year}{2008}\natexlab{}.
\newblock \showarticletitle{Block tridiagonal matrix inversion and fast transmission calculations}.
\newblock \bibinfo{journal}{\emph{J. Comput. Phys.}} \bibinfo{volume}{227}, \bibinfo{number}{6} (\bibinfo{date}{March} \bibinfo{year}{2008}), \bibinfo{pages}{3174--3190}.
\newblock
\showISSN{0021-9991}
\urldef\tempurl%
\url{https://doi.org/10.1016/j.jcp.2007.11.035}
\showDOI{\tempurl}


\bibitem[Polizzi and Sameh(2006)]%
        {pdiv_1}
\bibfield{author}{\bibinfo{person}{Eric Polizzi} {and} \bibinfo{person}{Ahmed~H. Sameh}.} \bibinfo{year}{2006}\natexlab{}.
\newblock \showarticletitle{A parallel hybrid banded system solver: the {SPIKE} algorithm}.
\newblock \bibinfo{journal}{\emph{Parallel Comput.}} \bibinfo{volume}{32}, \bibinfo{number}{2} (\bibinfo{date}{Feb.} \bibinfo{year}{2006}), \bibinfo{pages}{177--194}.
\newblock
\showISSN{0167-8191}
\urldef\tempurl%
\url{https://doi.org/10.1016/j.parco.2005.07.005}
\showDOI{\tempurl}


\bibitem[Rue et~al\mbox{.}(2009)]%
        {rue2009approximate}
\bibfield{author}{\bibinfo{person}{H{\aa}vard Rue}, \bibinfo{person}{Sara Martino}, {and} \bibinfo{person}{Nicolas Chopin}.} \bibinfo{year}{2009}\natexlab{}.
\newblock \showarticletitle{Approximate Bayesian inference for latent Gaussian models by using integrated nested Laplace approximations}.
\newblock \bibinfo{journal}{\emph{Journal of the Royal Statistical Society Series B: Statistical Methodology}} \bibinfo{volume}{71}, \bibinfo{number}{2} (\bibinfo{year}{2009}), \bibinfo{pages}{319--392}.
\newblock


\bibitem[Schenk and Gärtner(2004)]%
        {pardiso_1}
\bibfield{author}{\bibinfo{person}{Olaf Schenk} {and} \bibinfo{person}{Klaus Gärtner}.} \bibinfo{year}{2004}\natexlab{}.
\newblock \showarticletitle{Solving unsymmetric sparse systems of linear equations with {PARDISO}}.
\newblock \bibinfo{journal}{\emph{Future Generation Computer Systems}} \bibinfo{volume}{20}, \bibinfo{number}{3} (\bibinfo{date}{April} \bibinfo{year}{2004}), \bibinfo{pages}{475--487}.
\newblock
\showISSN{0167-739X}
\urldef\tempurl%
\url{https://doi.org/10.1016/j.future.2003.07.011}
\showDOI{\tempurl}


\bibitem[Schenk and Gärtner(2006)]%
        {pardiso_4}
\bibfield{author}{\bibinfo{person}{Olaf Schenk} {and} \bibinfo{person}{Klaus Gärtner}.} \bibinfo{year}{2006}\natexlab{}.
\newblock \showarticletitle{On fast factorization pivoting methods for sparse symmetric indefinite systems.}
\newblock \bibinfo{journal}{\emph{ETNA. Electronic Transactions on Numerical Analysis [electronic only]}}  \bibinfo{volume}{23} (\bibinfo{year}{2006}), \bibinfo{pages}{158--179}.
\newblock
\urldef\tempurl%
\url{http://eudml.org/doc/127439}
\showURL{%
\tempurl}


\bibitem[Schenk et~al\mbox{.}(2001)]%
        {pardiso_2}
\bibfield{author}{\bibinfo{person}{Olaf Schenk}, \bibinfo{person}{Klaus Gärtner}, \bibinfo{person}{Wolfgang Fichtner}, {and} \bibinfo{person}{Andreas Stricker}.} \bibinfo{year}{2001}\natexlab{}.
\newblock \showarticletitle{{PARDISO}: a high-performance serial and parallel sparse linear solver in semiconductor device simulation}.
\newblock \bibinfo{journal}{\emph{Future Generation Computer Systems}} \bibinfo{volume}{18}, \bibinfo{number}{1} (\bibinfo{date}{Sept.} \bibinfo{year}{2001}), \bibinfo{pages}{69--78}.
\newblock
\showISSN{0167-739X}
\urldef\tempurl%
\url{https://doi.org/10.1016/S0167-739X(00)00076-5}
\showDOI{\tempurl}


\bibitem[Sensi et~al\mbox{.}(2024)]%
        {nccl_alps}
\bibfield{author}{\bibinfo{person}{Daniele~De Sensi}, \bibinfo{person}{Lorenzo Pichetti}, \bibinfo{person}{Flavio Vella}, \bibinfo{person}{Tiziano~De Matteis}, \bibinfo{person}{Zebin Ren}, \bibinfo{person}{Luigi Fusco}, \bibinfo{person}{Matteo Turisini}, \bibinfo{person}{Daniele Cesarini}, \bibinfo{person}{Kurt Lust}, \bibinfo{person}{Animesh Trivedi}, \bibinfo{person}{Duncan Roweth}, \bibinfo{person}{Filippo Spiga}, \bibinfo{person}{Salvatore~Di Girolamo}, {and} \bibinfo{person}{Torsten Hoefler}.} \bibinfo{year}{2024}\natexlab{}.
\newblock \bibinfo{title}{Exploring {GPU}-to-{GPU} {Communication}: {Insights} into {Supercomputer} {Interconnects}}.
\newblock
\newblock
\urldef\tempurl%
\url{https://doi.org/10.48550/arXiv.2408.14090}
\showDOI{\tempurl}
\newblock
\shownote{arXiv:2408.14090}.


\bibitem[Spellacy and Golden(2018)]%
        {pdiv_4}
\bibfield{author}{\bibinfo{person}{Louise Spellacy} {and} \bibinfo{person}{Darach Golden}.} \bibinfo{year}{2018}\natexlab{}.
\newblock \showarticletitle{Partial {Inverses} of {Complex} {Block} {Tridiagonal} {Matrices}}. In \bibinfo{booktitle}{\emph{Parallel {Processing} and {Applied} {Mathematics}}}, \bibfield{editor}{\bibinfo{person}{Roman Wyrzykowski}, \bibinfo{person}{Jack Dongarra}, \bibinfo{person}{Ewa Deelman}, {and} \bibinfo{person}{Konrad Karczewski}} (Eds.). \bibinfo{publisher}{Springer International Publishing}, \bibinfo{address}{Cham}, \bibinfo{pages}{634--645}.
\newblock
\showISBNx{978-3-319-78024-5}
\urldef\tempurl%
\url{https://doi.org/10.1007/978-3-319-78024-5_55}
\showDOI{\tempurl}


\bibitem[Svizhenko et~al\mbox{.}(2002)]%
        {rgf_1}
\bibfield{author}{\bibinfo{person}{A. Svizhenko}, \bibinfo{person}{M.~P. Anantram}, \bibinfo{person}{T.~R. Govindan}, \bibinfo{person}{B. Biegel}, {and} \bibinfo{person}{R. Venugopal}.} \bibinfo{year}{2002}\natexlab{}.
\newblock \showarticletitle{Two-dimensional quantum mechanical modeling of nanotransistors}.
\newblock \bibinfo{journal}{\emph{Journal of Applied Physics}} \bibinfo{volume}{91}, \bibinfo{number}{4} (\bibinfo{date}{Feb.} \bibinfo{year}{2002}), \bibinfo{pages}{2343--2354}.
\newblock
\showISSN{0021-8979}
\urldef\tempurl%
\url{https://doi.org/10.1063/1.1432117}
\showDOI{\tempurl}


\bibitem[Venetis et~al\mbox{.}(2015)]%
        {pdiv_gpu}
\bibfield{author}{\bibinfo{person}{I.~E. Venetis}, \bibinfo{person}{A. Kouris}, \bibinfo{person}{A. Sobczyk}, \bibinfo{person}{E. Gallopoulos}, {and} \bibinfo{person}{A.~H. Sameh}.} \bibinfo{year}{2015}\natexlab{}.
\newblock \showarticletitle{A direct tridiagonal solver based on {Givens} rotations for {GPU} architectures}.
\newblock \bibinfo{journal}{\emph{Parallel Comput.}}  \bibinfo{volume}{49} (\bibinfo{date}{Nov.} \bibinfo{year}{2015}), \bibinfo{pages}{101--116}.
\newblock
\showISSN{0167-8191}
\urldef\tempurl%
\url{https://doi.org/10.1016/j.parco.2015.03.008}
\showDOI{\tempurl}


\bibitem[Woolley({[n.\,d.]})]%
        {nccl_2}
\bibfield{author}{\bibinfo{person}{Cliff Woolley}.} \bibinfo{year}{[n.\,d.]}\natexlab{}.
\newblock \showarticletitle{{NCCL}: {ACCELERATED} {MULTI}-{GPU} {COLLECTIVE} {COMMUNICATIONS}}.
\newblock  (\bibinfo{year}{[n.\,d.]}).
\newblock
\urldef\tempurl%
\url{https://images.nvidia.com/events/sc15/pdfs/NCCL-Woolley.pdf}
\showURL{%
\tempurl}


\bibitem[Zammit-Mangion and Rougier(2018)]%
        {zammit2018sparse}
\bibfield{author}{\bibinfo{person}{Andrew Zammit-Mangion} {and} \bibinfo{person}{Jonathan Rougier}.} \bibinfo{year}{2018}\natexlab{}.
\newblock \showarticletitle{A sparse linear algebra algorithm for fast computation of prediction variances with Gaussian Markov random fields}.
\newblock \bibinfo{journal}{\emph{Computational Statistics \& Data Analysis}}  \bibinfo{volume}{123} (\bibinfo{year}{2018}), \bibinfo{pages}{116--130}.
\newblock
\newblock
\shownote{\href{https://doi.org/10.1016/j.csda.2018.02.001}{10.1016/j.csda.2018.02.001}}.


\bibitem[Ziogas et~al\mbox{.}(2019)]%
        {rgf_gpu}
\bibfield{author}{\bibinfo{person}{Alexandros~Nikolaos Ziogas}, \bibinfo{person}{Tal Ben-Nun}, \bibinfo{person}{Guillermo~Indalecio Fern\'{a}ndez}, \bibinfo{person}{Timo Schneider}, \bibinfo{person}{Mathieu Luisier}, {and} \bibinfo{person}{Torsten Hoefler}.} \bibinfo{year}{2019}\natexlab{}.
\newblock \showarticletitle{A data-centric approach to extreme-scale ab initio dissipative quantum transport simulations}. In \bibinfo{booktitle}{\emph{Proceedings of the International Conference for High Performance Computing, Networking, Storage and Analysis}} (Denver, Colorado) \emph{(\bibinfo{series}{SC '19})}. \bibinfo{publisher}{Association for Computing Machinery}, \bibinfo{address}{New York, NY, USA}, Article \bibinfo{articleno}{1}, \bibinfo{numpages}{13}~pages.
\newblock
\showISBNx{9781450362290}
\urldef\tempurl%
\url{https://doi.org/10.1145/3295500.3357156}
\showDOI{\tempurl}


\end{thebibliography}

\end{document}